\documentstyle[12pt,aps,epsfig,prb]{revtex}

\def\bi{\bibitem}

\newcommand{\unoint}{\int \frac{d^{d} q}{(2 \pi)^{d}}}

\newcommand{\bmat}{\begin{mathletters}}
\newcommand{\emat}{\end{mathletters}}
\newcommand{\be}{\begin{equation}}
\newcommand{\ee}{\end{equation}}
\newcommand{\bq}{\begin{eqnarray}}
\newcommand{\eq}{\end{eqnarray}}

\newcommand{\nn}{\nonumber}

\def\eps{\epsilon}

\def\ff{\tilde{f}}
\def\rff{\rho \tilde{f}}
\def\f0{\tilde{f}(0)}
\def\rf0{\rho \tilde{f}(0)}
\def\ep{\epsilon(p)}

\newcommand{\diffe}[1]{\left[ \rff(p-\emph{#1})-\rff(\emph{#1}) \right]}

\begin{document}

\title{The dynamical structure factor in topologically
disordered systems}
\draft
\address{Victor Martin-Mayor$^{a}$,
Marc M\'ezard$^{b}$, Giorgio Parisi$^{a}$ and Paolo Verrocchio$^{c}$}
\address{a) Dipartimento di Fisica, Sezione INFN and Unit\`a INFM,\\
Universit\`a di Roma ``La Sapienza'',
Piazzale Aldo Moro 2,
I-00185 Rome (Italy)}
\address{b)  Laboratoire de Physique Th\'eorique et Mod\`eles Statistiques,
Universit\'e de Paris Sud, Bat.100, 91405 Orsay
(France) \\}
\address{c) Dipartimento di Fisica, Universita' di Trento, Via 
Sommarive 14, I-38050 Povo, Trento (Italy)}
\date{\today}

\maketitle

\begin{abstract}
A computation of the dynamical structure factor of topologically
disordered systems, where the disorder can be described in terms of
euclidean random matrices, is presented. Among  others, structural glasses and
supercooled liquids belong to that class of systems.  The computation
describes
their relevant spectral features in the region of 
the high frequency sound.  
The analytical results
are tested with numerical simulations and are found to be in very good
agreement with them.  Our results may explain the findings of
inelastic X-ray scattering experiments in various glassy systems.
\end{abstract}

\section{Introduction}

\noindent
Inelastic X-ray scattering (IXS)
experiments~\cite{Ben,Mas,Mas2,MoMaRuSe,Mas3,Fio,Sok,Ruo,Set} and
inelastic neutron scattering~\cite{Buc,FoCoVaSu} on structural glasses
and supercooled liquids provided useful information on the dynamics of
their amorphous structure, at frequencies larger than $0.1$ THz.
Those experiments show a regime, when the wavelength of the plane wave
is comparable with the inter-particle distance, where the vibrational
spectrum can be understood in terms of propagation of quasi-elastic
sound waves, the so-called high frequency sound. This high-frequency
sound has also been observed in molecular dynamical simulations of
strong~\cite{HoKoBi1,TaEl,DeRuSaVi,FeAllBi,AllFeFaWo,RiWiMa} and
fragile~\cite{MaRuSa,Sam} liquids, and it displays
several rather universal
features~\cite{Mas,Mas2,MoMaRuSe,Mas3,Fio,Sok,Ruo,Set,HoKoBi1,DeRuSaVi,TaEl,DeRuSaVi,FeAllBi,AllFeFaWo,RiWiMa,MaRuSa,Sam}
(see however Ref.~\cite{FoCoVaSu} for a dissenting view regarding the
silica properties). In particular, a peak is observed in the dynamical
structure factor at a frequency which depends linerly on
 the exchanged momentum $p$,
in the region $0.1p_0-1.0p_0$, $p_0$ being the position of the first maximum
in the static structure factor. When extrapolated to zero momentum, 
this linear dispersion relation yields the macroscopic speed of sound.
The width of the spectral line, $\Gamma$ is well fitted by
\begin{equation}
\Gamma(p)= A p^x\quad ,\quad x\approx 2\,,
\label{BROADENING}
\end{equation}
with $A$ displaying a very mild (if any) temperature
dependence. Although the same scaling of the spectral line is found in
hydrodynamics~\cite{Ha}, in this computation the proportionality 
constant is basically the
viscosity, which shows a very strong temperature dependence which is not observed.
Moreover, the same scaling of $\Gamma$ has been found in harmonic
Lennard-Jones glasses~\cite{Ruo2}, and one can safely conclude
that the $p^2$ broadening of the high-frequency sound has a physical
origin different from hydrodynamics. Other interesting problems
related with the high-frequency vibrational excitations of these
topologically disordered systems~\cite{Ell} regard the origin of 
the {\em Boson
peak} or the importance of localization properties to understand the
dynamics of supercooled liquids~\cite{BeLa}.

The variety of materials in which
the $p^2$ broadening appears suggests a straightforward physical
motivation. However, the simplest conceivable approximation, a wave
propagating on an elastic medium in the presence of random scatterers,
yields Rayleigh dispersion: $\Gamma\propto p^4$. This result is
very robust: as soon as one assumes the presence of an underlying medium on
which the sound waves would propagate undisturbed, as in the
disordered-solid model~\cite{ScDiGa,Montagna}, the $p^4$ scaling
appears even if one studies the interaction with the scatterers
non-perturbatively~\cite{MaPaVe}. In this paper we want to show that
when the distinction between the propagating medium and the scatterers
is meaningless (as it happens for topologically disordered systems),
the $p^2$ scaling is recovered. 
Recently, it has been shown~\cite{GoMa} that the numerical solution of
the mode coupling equations, modified for the study of the glassy
phase, describes in a qualitatively correct way the range of
frequencies explored by IXS scattering, including the high frequency sound
and the  boson peak.  In that theoretical approach though, one
cannot find a clear scaling law, $\Gamma$ being proportional to $p^2$
only at very low momentum and following a more complicated law for
higher values. 

We want to investigate the problem from the point of view of
statistical mechanics of random matrices~\cite{Metha}, by assuming
that vibrations are the only motions allowed in the system.  The
formalism we shall introduce, however, is not limited to the
investigation of the high frequency sound and it could be straightforwardly
applied in different physical contexts.


Let us look more carefully at the relation between the
high frequency behaviour and vibrational dynamics in glasses
and at the relation between vibrations and random matrices.
The dynamical structure factor
for a system of $N$ identical particles is defined as~\cite{Ha}:
\begin{equation}
S(p,\omega) = \frac{1}{N} \sum_{i,j} \int dt\ 
 {\mathrm e}^{{\mathrm i}\omega t}\ \left\langle\, {\mathrm e}^{{\mathrm i}  p
\cdot\left( r_j(t)- r_i(0) \right)}\, \right\rangle\, ,
\label{FOURIER}
\end{equation}
where $\langle \dots \rangle$ denotes  the average over the particles
positions $r_j$ with the canonical ensemble.

A well known approach to the calculation of the $S(p,\omega)$ in the
high-frequency (THz) region, consists in taking into account only the
vibrational modes of the system. The basic assumption of this
calculational strategy is that there is a clear separation among the
time scales of the 'fast' degrees of freedom, which are supposedly
modeled as vibrations, and the 'slow' ones, which give rise to the
diffusion of the particles. One expects this assumption to be
reasonably accurate in the glass and in the super-cooled liquid phase.

When focusing only on the short time region, the slow motions can be
thought as 'frozen', hence 
one studies the displacements $u$ around 'quenched' positions $x$, by
writing the position of the $i$-th particle as ${r}_i(t)=
{x}_i + {u}_i(t)$,
and linearizing the equations of motion. Then one is naturally lead
to consider the spectrum of the Hessian matrix of the potential, evaluated
on the positions $x$ of the particles.
Calling $\omega_n^2$  the eigenvalues of the Hessian matrix
and $ e_n(i)$ the corresponding eigenvectors, 
the one excitation approximation to the $S(p,\omega)$
at non zero frequency is 
given  in the classical
limit by:
\begin{eqnarray}
S^{(1)}(p,\omega)&=&\frac{k_{\mathrm B}T}{m \omega^2}\,\sum_{n=1}^N\, 
Q_n(p)\,\delta(\omega-\omega_n)\,,\\
Q_n(p)&=&
|\sum_i\, 
 p\cdot e_n(i) {\mathrm e}^{{\mathrm i} p\cdot x_i} |^2\,.
\label{ONE}
\end{eqnarray}

However, in the supercooled liquid phase, one cannot always assume that
in a typical configuration at equilibrium all the normal modes have positive 
eigenvalues, negative eigenvalues 
representing the situation where some particles are moving away from the 
position $x$. 
This observation has motivated the so-called
instantaneous normal modes (INM) approach~\cite{Ke,WuSt,WuLo,CaGiPa}
in the study of supercooled liquids. In that approach 
the normal modes in a typical configuration are 
supposed to describe the short time dynamics of the particles in the 
liquid phase, while it
has been suggested~\cite{BeLa} that diffusion properties 
could be studied by considering the localization
properties of the normal modes of negative eigenvalues.

On the other hand, it has been argued that in a very broad class of glassy 
systems below a critical temperature~\cite{KiThWo}, 
called dynamical temperarture $T_{D}$ (for spin glasses) or mode coupling 
temperature $T_{MCT}$ (for structural glasses), the particles spend the
most of their time around the minima of the potential energy 
surface. Recent numerical simulations seem to confirm that
picture~\cite{BrBaCaZiGi,AnDiRuScSc}.

As a consequence, in the glassy phase the high frequency window
($\geq .1$ Thz) of the dynamical structure factor is supposed to be
well described by the propagation of quasi-harmonic excitations around
a {\em quenched} structure \cite{GoMa}.  
By assuming that this structure corresponds to one of the many minima 
of the potential energy, 
one can introduce a harmonic approximation where
only the vibrations around these minima are considered, and all the
dynamical information is encoded in the spectral properties of the
Hessian matrix evalutated on the rest positions, where it has not
negative eigenvalues.
It has been shown using molecular
dynamics simulation that below the experimental glass transition
temperature the thermodynamical properties of typical strong glasses
\cite{HoKoBi2} are in a good agreement with such an assumption.
Let us note that, within this assumption, the dynamical structure 
factor (\ref{ONE}) depend on the minimum around which the
particles oscillate, hence the knowledge of an infinite number of
disordered positions $x$ seems to be needed.
Here we shall make the  reasonable assumption that
the dynamical structure factor is self-averaging, a type of
assumption which  generally turns out to 
be correct for macroscopic observables in disordered systems.

Hence we have turned our  problem into the  study
of  spectral properties 
of a kind of random matrix, once the probability distribution of the 
elements of the matrix is given.
More precisely, the problem of the high-frequency dynamics of the system
can be reduced, in its simplest version,
to the consideration of a very particular type of
random matrices, the euclidean matrices~\cite{MePaZe}, in which the
entries are deterministic functions (the derivatives of the
potential) of the random positions of the
particles. If the system has a conservation law such as momentum
conservation in our case, due to translational invariance, all the
realizations of the random matrix have a common normal mode with zero
eigenvalue: the uniform translation of the system.  An euclidean matrix
is determined by the particular deterministic function that we are
considering, and by the probabilistic distribution function of the
particles. 
However, for the sake of simplicity we shall concentrate here
on the simplest kind of euclidean matrices~\cite{MePaZe}. We consider
$N$ particles placed at positions $x_i, i=1,2,...,N$ inside a box of volume 
$V$, where
periodic boundary conditions are applied. Then, the scalar euclidean matrices
are
\be
M_{ij} \equiv \delta_{ij} \sum_{k=1}^N f( x_i - x_k) 
- f( x_i -x_j)\quad ,
\quad i,j=1,2\ldots,N\ .
\label{EUCLIDEA}
\ee 
In the above expression, $f(x)$ is an arbitrary but
deterministic scalar function depending on the distance between pairs
of particles, and the positions $\{ {x} \}$ of the particles are
drawn by a probabilistic law $P[ {x} ]$.  Notice that the matrix
(\ref{EUCLIDEA}) preserves translation invariance, since the uniform
vector $e_0(i)=const$ is an eigenvector of $M$ with  zero eigenvalue. 
Let us note furthermore that there are not internal indices, since for 
the sake of simplicity the particle displacements are restricted to
be all collinear. Of course, in the study of the vibrations of glasses, 
one should take into account the vectorial nature of those vibrations and 
split the Hessian matrix in its transversal and longitudinal 
(with respect to the directions of the external momentum $\vec{p}$) parts. 
A careful analysis of that feature will be presented
elsewhere \cite{futuro}.

Simple as these matrices are, many physical problems
ranging from vibrational spectra in amorphous systems and electron
hopping in amorphous semiconductors \cite{Ell}, to instantaneous
normal modes (INM) in liquids \cite{Ke,WuSt,WuLo,CaGiPa} and
combinatorial optimization \cite{MePa}, can be described with them
using an appropriate choice of $P$.  In this paper we will study the
simplest case, in which $P$ is a uniform distribution function.

The dynamical structure factor for a scalar euclidean-matrix is given
by
\begin{eqnarray}
S_E(p,E)&=& \overline{\sum_n Q_n(p)\delta(E - E_n)}\,,\label{DEFINIZIONE}\\
Q_n(p)&=& \frac{1}{N}\left|\sum_{i=1}^N\, e_n(i) 
{\mathrm e}^{{\mathrm i}  p\cdot x_i}\right|^2\,,\\
S(p,\omega)&=&2\omega S_E(p,\omega^2)\,,
\label{NORMA}
\end{eqnarray}
where the overline stands for the average over the particles position and 
we have given the definition either in the eigenvalue space ($S_E(p,E)$)
and in the frequency space ($S(p,\omega)$). 

Finally, let us mention that not all the topologically disordered
systems are represented by an euclidean  random matrix. For instance,
diffusion on random graphs \cite{BrRo,BiMo} and on small-world network
\cite{Mo} where the connecting sites are chosen with a probabilistic
law, are described by the Laplacian operator ${\cal W}_{ij}$. The
matrix element ${\cal W}_{ij}$, $i \neq j$ is one if $i$ and $j$ are
connected and $0$ otherwise, while the diagonal part is tailored to
have $\sum_i {\cal W}_{ij} =0$, as required by the conservation of the
number of diffused particles.


The layout of the rest of this paper is organized as follows. In
section~\ref{SVILUPPOSECT} we shall apply an expansion in the inverse
of the particle density formulation to the problem of calculating the
$S_E(p,E)$. The same calculation can be performed using the field
theory of Ref.~\cite{MePaZe}, adapted to this problem~\cite{NOSOTROS},
with identical results~\cite{NOSOTROS}.  The analytical computations
will be confronted with a numerical simulation in
section~\ref{ONELOOPSECTNUM}.  In section~\ref{CORRELATIONS} we shall
briefly consider the problem of including particle correlations on our
calculation, which shall produce some qualitatively new results,
regarding the exponential tail of the density of states. In
section~\ref{CONCLUSIONS} we shall present our conclusions.

\section{The 1/$\rho$ expansion}\label{SVILUPPOSECT}

The basic object that we will
calculate is the resolvent
\bq
G(p,z) &=& {1 \over N}
\overline{\sum_{ij} e^{i p(x_i-x_j)} 
\left[\frac{1}{z - M}\right]_{ij} }\ ,
\label{GDIPI}
\eq
from which the dynamical structure factor is recovered using the
distribution identity \hbox{$1/(x+i0^{+})=P(1/x)-{\mathrm i}\pi\delta(x)$}:
\begin{equation}
S_E(p,E)=-\frac{1}{\pi}\lim_{\eta\to 0^+}{\mathrm Im}\ G(p,E+{\mathrm i}\eta)\,.
\end{equation}
The resolvent will also give us access to the density of states of
the system:
\begin{eqnarray}
g_E(E)&=&
-\frac{1}{N\pi}\lim_{\eta\to 0^+} {\mathrm Im}\  \overline{\sum_{i=1}^N 
\left[\frac{1}{E+{\mathrm i}\eta - M}\right]_{ii}}\,,\\
&=&-\frac{1}{\pi}\lim_{\eta\to 0^+}{\mathrm Im}\, \lim_{p\to\infty}G(p,E+{\mathrm i}\eta)\,.
\label{DOF}
\end{eqnarray}

We shall set up a perturbative expansion in the inverse density of
particles, $1/\rho$. The zero$^{th}$ order of this expansion (to be
calculated in subsection~\ref{ZEROSECT}) corresponds to the limit of
infinite density, in which the system is equivalent to an elastic
medium.  In this limit the resolvent  turns out to be extremely
simple:
\begin{eqnarray}
G(p,z)&=&\frac{1}{z-\epsilon(p)}\,,
\label{ZEROTH}\\
\ep&=&\rho[\ff(0)-\ff(p)]\label{LAEDIPI}\,.
\end{eqnarray}
In the above expression $\ff(p)$ is the Fourier transform of the function $f$
defined in Eq.(\ref{EUCLIDEA}), that due to its spherical symmetry, 
behaves like $\ep={\cal O}(p^2)$ for small $p$.
We see that the dynamical structure function has two delta functions at 
frequencies linear in $p$:
\hbox{$\omega=\pm\sqrt{\epsilon(p)}$}.  It is then clear that
Eq.(\ref{ZEROTH}) represents the undamped propagation of sound in our
elastic medium, with a dispersion relation controlled by the function
$\ep$ . 
The order $1/\rho$ corrections to  Eq.(\ref{ZEROTH}) will
be calculated in subsection~\ref{ONELOOPSECT}. They take the form of a
complex self-energy, $\Sigma(p,z)$, yielding
\begin{eqnarray}
G(p,z) &=&  \frac{1}{z-\ep -\Sigma(p,z)}
\label{CORRETTO}\,,\\
S_E(p,E)&=&  -\frac{1}{\pi} \frac{Im \Sigma(p,E)}
{\left(E-\ep -Re \Sigma(p,E)\right)^2 +\left(Im \Sigma(p,E)\right)^2 }
\label{ALLARGAMENTO}\,.
\end{eqnarray}
The dynamical structure factor,
is no longer a delta function, but
close to its maxima it has a Lorentzian shape. From
Eq.(\ref{ALLARGAMENTO}) we see that the real part of the self-energy
renormalizes the dispersion relation, and thus
it should behaves proportionallyto $p^2$ at small momentum. 
The width of the
spectral line is instead controlled by the imaginary part of the
self-energy, and at this order in $1/\rho$, it scales like $p^4$ for
small momentum: at this level of accuracy we recover Rayleigh
scattering. The analytic
calculation to this order can be confronted with numerical simulations
(see section~\ref{ONELOOPSECTNUM}), and it turns-out to be rather
accurate for momenta of the order of the inverse interparticle
distance. The density of states for small values of $\omega$ is also
well reproduced. The maybe most interesting result of this paper is
obtained in section~\ref{TWOLOOPSECT}, where the calculation of the
self-energy is pushed up to order $1/\rho^2$.  The order $1/\rho^2$
contribution to the spectral line-width is proportional to $p^2$ for
small momentum, which is a qualitatively new feature. Even if the
density of the system is high, for small-enough momenta the two loops
contribution to the imaginary part of the self-energy is
dominating. We believe this to be the underlying physical reason for
the widespread experimental finding of $p^2$ scaling of the width of
the dynamical structure factor. On the other hand, for larger momenta
the order $1/\rho$ contribution is dominating, but it grows with $p$
significantly more slowly than $p^4$. This maybe of some importance in
the experimentally relevant momentum range.

\subsection{The leading order}\label{ZEROSECT}

Our strategy for computing the resolvent (\ref{GDIPI}) is
straightforward: we consider it as the sum of a geometric series. One
expands in series of $1/z$ the Fourier transform of the resolvent, see
Eq.(\ref{GDIPI}), in the following way: 
\be G(p,z) = \frac{1}{z} \sum_R \left( \frac{-1}{z} \right)^R
\overline{M^R(p)}
\label{DIRETTA}
\ee
with
\bq
M^R (p) &=& \frac{1}{N} \sum_{k_0,k_1 \ldots k_R} \: e^{ip x_{k_0}} \left( \delta_{k_0, k_1} 
\sum_{z_1} f(x_{k_0} - x_{z_1}) - f(x_{k_0} -x_{k_1}) \right) \dots \nn \\ &\dots& \left(
\delta_{k_{R-1}, k_R} 
\sum_{z_R} f(x_{k_{R-1}} - x_{z_N}) - f(x_{k_{R-1}}-x_{k_R})\right )
e^{-ip x_{k_R}}
\label{GEO}
\eq 
Since we are interested in the singularities (branch-cuts or
poles) on the complex $z$ plane of the resolvent $G(p,z)$, see
Eqs.(\ref{ZEROTH},\ref{CORRETTO}), it will be crucial to sum the
series in Eq.(\ref{GEO}) at all orders in $1/z$. From now on we shall
assume that the function $f$ has a Fourier transform (denoted by $\ff$ in
the following) which is a rather severe limitation. In
section~\ref{CORRELATIONS} we will see that, provided that particle
correlations are taken into account on the calculation, this
limitation should not bother us.  One can readily check from
Eq.(\ref{GEO}) that the contribution of the terms where the two
particle labels in the argument of any $f$ function are the same,
vanish. This can be taken care of automatically by setting
$f(0)=0$. Since we have changed the value of the function at a single
point, this will have no consequences while averaging over the
positions of different particles.

The very first step is to split the parenthesis in Eq.(\ref{GEO}), into $2^R$ sums
(we keep the Kronecker deltas as a reminder of which $k_i$ index should
be equal to which):
\begin{eqnarray}
M^R (p) &=&\sum_{k_0,z_0,\ldots,z_{R-1}}\: 
\frac{e^{ip (x_{k_0}-x_{k_R})}}{N}\,[\delta_{k_0,k_1} f(x_{k_0}-x_{z_0})]
[\delta_{k_1,k_2} f(x_{k_1}-x_{z_1})]\ldots\nn\\\nonumber
&&\ldots[\delta_{k_{R-1},k_R} f(x_{k_{R-1}}-x_{z_{R-1}})]+\\\nonumber
&+&\sum_{k_0,k_1,z_1,\ldots,z_{R-1}}\: 
\frac{e^{ip (x_{k_0}-x_{k_R})}}{N}\,[- f(x_{k_0}-x_{k_1})]
[\delta_{k_1,k_2} f(x_{k_1}-x_{z_1})]\ldots\nn\\\nonumber
&&\ldots[\delta_{k_{R-1},k_R} f(x_{k_{R-1}}-x_{z_{R-1}})]+\\\nonumber
&& \ldots\\
&+&\sum_{k_0,k_1,\ldots,k_R}\: 
\frac{e^{ip (x_{k_0}-x_{k_R})}}{N}\,[- f(x_{k_0}-x_{k_1})]
[-f(x_{k_1}-x_{k_2})]\ldots[-f(x_{k_{R-1}}-x_{z_{R-1}})]
\label{SINPARENTESIS}
\end{eqnarray}
The disorder average in Eq.(\ref{SINPARENTESIS}) is simply obtained by
performing the integrals:
\begin{equation}
\overline{M^R(p)}=\int \prod_j \frac{d^D x_j}{V}\ M^R(p)\,
\end{equation}
Let us estimate the order of magnitude of the generic term in
Eq.(\ref{SINPARENTESIS}), after the disorder average is performed. If
there are no index repetitions, we will have an overall $V^{-R}$
factor from the disorder average, because although there are $R+1$
independent particles the translational invariance allows one to
eliminate one integral. We have $N!/(N-R-1)!\sim N^{R+1}$ ways of
choosing the particles labelling, and an extra $N^{-1}$ factor from
the definition of the resolvent (\ref{GDIPI}). Therefore, each of the
above $2^R$ sums will be proportional to $\rho^R$ ($\rho=N/V$ is the
particle density). When two particle label coincide, we will have a
missing $N$ factor from the sums, and another missing $V^{-1}$ factor
from the position average, and the sum will be of order
$\rho^{R-1}$. It is thus clear that the $\overline{M^R(p)}$ will be a
polynomial of order $R$ in $\rho$, with no term of order $\rho^0$. In
fact, the lowest order in $\rho$ arise when there are only two
different particle labels on the sums of Eq.(\ref{SINPARENTESIS}),
which is easily seen to be linear in $\rho$:
\begin{equation}
\overline{M^R(p)}= \rho^R\,{\cal I}^{(R)}_R(p)\; +\;
\rho^{R-1}\,{\cal I}^{(R)}_{R-1}(p)\; +\; \ldots\; +\;
\rho\,{\cal I}^{(R)}_{1}(p)\, .
\label{ILSUCO}
\end{equation}
Since $M^R(p=0)=0$, for {\em every} particle configuration, all the
coefficient ${\cal I}^{(R)}_{k}(p)$ in Eq.(\ref{ILSUCO}) vanish at zero
external momentum, $p=0$.

Our zero$^{th}$-order approximation will consist in keeping only the leading
term
\begin{equation}
\overline{M^R(p)}\approx \rho^R\,{\cal I}^{(R)}_R(p)\,,
\end{equation}
the first $1/\rho$ correction will consist in setting
\begin{equation}
\overline{M^R(p)}\approx \rho^R\,{\cal I}^{(R)}_R(p)\; +\;
\rho^{R-1}\,{\cal I}^{(R)}_{R-1}(p)\;\,,
\end{equation}
and so on.

Let us now calculate ${\cal I}^{(R)}_R(p)$.
As for any multidimensional integral, choosing with care the order in
which the integrations are performed can tremendously simplify the
task.  In the case of Eq.(\ref{SINPARENTESIS}) the matrix product
yields a fairly natural ordering: from left to right, and when we
encounter a diagonal term ($\delta_{k_i,k_{i+1}}$) we shall first
average the position of the $z_i$ particle. One easily shows that
(recall that the Kronecker delta is simply a {\em reminder}, since
$k_i$ is actually equal to $k_{i+1}$):
\begin{eqnarray}
\int dx_{z_i} e^{ip x_{k_i}} \delta_{k_i,k_{i+1}}f(x_{k_i}-x_{z_i})&=&
e^{ip x_{k_{i+1}}} \ff(0)\,,\label{SHIFTFORWARD}\\
\int dx_{k_i} e^{ip x_{k_i}}
[-f(x_{k_i}-x_{k_{i+1}})]&=& e^{ip x_{k_{i+1}}} [-\ff(p)]\,.\nonumber 
\end{eqnarray}
Therefore every integral simply shifts the external moment, $p$, one
step forward, and leaves behind a $\ff(0)$ or a $-\ff(p)$. When one
arrives to the ${R+1}$ particle, there is no integral left to be done, and both
exponentials cancel. Then we find that
\begin{equation}
\rho^R{\cal I}^{(R)}_R(p)=\left[\rho\ff(0)-\rho\ff(p)\right]^R\,,
\end{equation}
and therefore we obtain
the result already anticipated in Eq.(\ref{ZEROTH})
\begin{equation}
G_0(p,z) = \frac{1}{z-\ep}\,.
\end{equation}
We see that at  leading order, a plane wave with momentum $p$ is
actually an eigenstate of the matrix $M$ with eigenvalue $E$, and the
disorder does not play any relevant role. In other words, inside a
wavelength $2 \pi /p$ there is always an infinite number of particles,
smoothing out the density fluctuations of the particles: the system
reacts as an elastic medium.

Let us finally obtain the density of states at this level of accuracy, using
Eq(\ref{DOF}) and $G_0$:
\begin{equation}
g_E(E)=\delta\left(E-\rho\ff(0)\right)\,.\label{EINSTEIN}
\end{equation}
We obtain a single delta function at $\rho\ff(0)$, which is somehow
contradictory with our result for the dynamical structure factor: from
the density of states one would say that the dispersion relation is
Einstein's like, without any momentum dependence! The way out of this
contradiction is of course that in the limit of infinite $\rho$ both
$\ep$ and $\rho\ff(0)$ diverge. 
The delta function in eq. (\ref{EINSTEIN}) is the leading term in $\rho$, 
while the states 
which contribute to the dynamical  structure factor
appear only in the subleading terms in the 
density of states.

\subsection{One loop}\label{ONELOOPSECT}

For the calculation of the ${\cal I}^{(R)}_{R-1}(p)$ term, we need to
consider only one particle label repetition. In other words, if we call
$N_R$  the average number of particles within the interaction range
of the potential, we are calculating the 1/$N_R$ correction of the
formula $\overline{M^R}=\overline{M}^R$, which neglects the statistical
fluctuation of the matrix $M$ due to disorder.

The easiest way to proceed is as follows: one first identifies the places
where the particle label repetition arises in the sequence of particle labels
$$(\ldots 1\ldots 1\ldots)\,,$$ and then one shifts the moment using
Eqs.(\ref{SHIFTFORWARD}) until the place of the first repeated
particle label. Then, one applies the backward version of the moment
shift idea, from $k_{R}$ until the second particle label repetition,
and one finally integrates over the position of the repeated particle.
In this way one gets:
\begin{eqnarray}
\frac{\rho^{R-1}\,{\cal I}^{(R)}_{R-1}(p)}{z^{R+1}}&=& \sum_{a+b+c=R-2} 
\frac{[\rho(\ff(0)-\ff(p))]^{a}}{z^{a+1}}
\frac{[\rho(\ff(0)-\ff(p))]^{c}}{z^{c+1}}\times\\\nonumber
&\times&
\int\frac{dq}{\rho(2\pi)^D} [\rho(\ff(p-q)-\ff(q))]^2
\frac{[\rho(\ff(0)-\ff(q))]^b}{z^{b+1}}\,,
\label{UNOSOBRERHOMOMENTO}
\end{eqnarray}
which is easily seen to yield
\begin{equation}
\sum_{R=0}^\infty\,\frac{\rho^{R-1}\,{\cal I}^{(R)}_{R-1}(p)}{z^{R+1}}= 
[G_0(z,p)]^2\,\frac{1}{\rho}
\int\frac{dq}{(2\pi)^D} G_0(q)\,[\rho(\ff(p-q)-\ff(q))]^2 \,. 
\label{UNOSOBRERHO}
\end{equation}

We want  to obtain the self-energy term in
(\ref{CORRETTO}), which will receive contributions of all
powers in $1/\rho$:
\begin{equation}
\Sigma(p,z)=\Sigma_1(p,z)+\Sigma_2(p,z)+\ldots
\end{equation}
The relation between the above expresion and the resolvent is given by
the  Dyson resummation:
\begin{equation}
G(p,z)-G_0(p,z)=G_0(p,z)\left[\Sigma_1(p,z)+ G_0(p,z)\Sigma^2_1(p,z) + 
\Sigma_2(p,z)+\ldots\right]G_0(p,z)\,.
\label{DYSONSUMA}
\end{equation}
Comparing Eqs.(\ref{DYSONSUMA}) and (\ref{UNOSOBRERHO}), we conclude that
\be
\Sigma_1(p,z) \equiv \frac{1}{\rho}\unoint \: G_0(q,z) \diffe{q}^2 \ .
\label{SELF1}
\ee 
The $1/\rho^2$ contributions to the self-energy will provide the
$G^3_0(p,z)\Sigma^2_1(p,z)$ term of Eq.(\ref{DYSONSUMA}), and also the
more physically interesting $\Sigma_2(p,z)$ term.

Let us study in details the low exchanged momentum limit of 
Eq.(\ref{SELF1}). It is clear that at $p=0$ the self-energy vanishes,
as required by the translational invariance. 
We need to expand $\ff(p-q)$ for small $p$, which due to the spherical
symmetry of $\ff$ yields
\begin{eqnarray}
\ff(p-q)&=&\ff(q)- ( p\cdot  q)\, \frac{\ff'(q)}{q}+{\cal O}(p^2)\,,\\
&=&\ff(q)+( p\cdot q)\,  \frac{\epsilon'(q)}{q\rho}+{\cal O}(p^2)\,,
\label{PICCOLOP}
\end{eqnarray}
where we have used Eq.(\ref{LAEDIPI}) for the derivative of $\ff$. 
Substituting (\ref{PICCOLOP}) in (\ref{SELF1}), and performing
explicitly  the trivial angular integrations in dimensions $d$ we obtain
\begin{eqnarray}
\Sigma_1(p,z)&\approx& p^2 \frac{2^{1-d}}{\rho d \pi^{d/2}\Gamma(d/2)}
\int_0^\infty dq\,q^{d-1}\frac{\left[\epsilon'(q)\right]^2}{z-\epsilon(q)}\,,\\
&=&p^2 \frac{2^{1-d}}{\rho d \pi^{d/2}\Gamma(d/2)}
\int_0^{\epsilon(q=\infty)} d\epsilon\,
\frac{\left[q(\epsilon)\right]^{d-1}}{q'(\epsilon)(z-\epsilon)}\,.\\
\end{eqnarray}
In the last equation, we have denoted with $q(\epsilon)$ the inverse
of the function $\epsilon(q)$.  Setting now $z= E+{\mathrm i} 0^+$,
and observing that $\ep\approx A p^2$ for small $p$, we readily obtain
\begin{eqnarray}
Re \Sigma_1(p, E+{\mathrm i} 0^+)&\approx& 
p^2 \frac{2^{1-d}}{\rho d \pi^{d/2}\Gamma(d/2)} P
\int_0^{\epsilon(q=\infty)} d\epsilon\,
\frac{\left[q(\epsilon)\right]^{d-1}}{q'(\epsilon)(E-\epsilon)}\,,
\label{REASSI1L}
\\
Im \Sigma_1(p, E+{\mathrm i} 0^+)&\approx& -
 \frac{\pi 2^{2-d} A}{\rho d \pi^{d/2}\Gamma(d/2)} p^2 [q(E)]^d\,.
\label{IMASSI1L}
\end{eqnarray}
Since the principal part is a number of order one,
the real part of the self-energy  scales like $p^2$ (possibly
with logarithmic corrections), and thus the speed of sound of the
system renormalizes due to the $1/\rho$ corrections. As a consequence,
the function $q(E)$ is proportional to 
$E^{1/2}\sim p$ at the maximum of the function of $p$
$S_E(p,E)$, and the width of the peak of the $S_E(p,E)$ will scale like
$p^{d+2}$. It is then easy to check (see (\ref{NORMA}))
 that in frequency space the width
of the spectral line will scale like 
\be 
\Gamma \propto p^{d+1}\,,
\label{FREQUENZE}
\ee
as one would expect from Rayleigh scattering considerations. 
The result (\ref{FREQUENZE})  for  the asymptotic regime $p<<1$
has been found at the one loop level.
In order to predict correctly the spectral properties at very low external 
momentum $p$, it turns out that one must  study the behaviour of the two loop
contribution, as we shall see in the next section. Nevertheless,
the one loop result is already a good starting point to perform detailed comparisons 
with the numerical simulations. To this end it will be useful in the following to 
introduce a concrete example with a specific choice of the function
$f(x)$, allowing a complete analytical computation.

\subsubsection{The Gaussian case}

Our choice for the function $f$ shall be the following
\begin{eqnarray} 
f(x)&=&{\mathrm e}^{-\frac{x^2}{2\sigma^2}}\,,\label{SCELTA}\\
\ff(p)&=&(2\pi\sigma^2)^{d/2}\,{\mathrm e}^
{-\sigma^2\frac{p^2}{2}}\, \label{SCELTAQ}.
\end{eqnarray}
The parameter $\sigma$ sets our length-scale, and for phenomenological
purposes might be identified with the inverse of the first maximum of
the static structure factor. With this choice, we have
\begin{eqnarray}
\epsilon_{max}&\equiv&\lim_{q\to\infty}\epsilon(q)=
\rho (2\pi\sigma^2)^{d/2}\,, \nn \\
\epsilon(q)&=&\epsilon_{max}
\left(1-{\mathrm e}^{-\sigma^2\frac{q^2}{2}}\right)\,,\\
q(\epsilon)&=&\frac{1}{\sigma}\sqrt{2
\ln{\frac{\epsilon_{max}}{\epsilon_{max}-\epsilon}}}\nn \,.
\end{eqnarray}
It is then straightforward to obtain in three dimensions
\begin{eqnarray}
\Sigma_1(p,z)&=&\sqrt{\frac{2\sigma^2}{\pi}}\int_0^{\epsilon_{max}}\,
d\epsilon\,\frac{{\cal G}\left(q(\epsilon),p\right)}{z-\epsilon}\,,\\
{\cal G}(q,p)&=&q\, {\mathrm e}^{-\sigma^2\frac{q^2}{2}}
\left[1+{\mathrm e}^{-\sigma^2 p^2}\frac{{\mathrm{sh}}(2\sigma^2 p q)}{2\sigma^2 p q}-2 
{\mathrm e}^{-\sigma^2 \frac{p^2}{2}}\frac{{\mathrm{sh}}(\sigma^2 p q)}{\sigma^2 p q} 
\right] \nn
\end{eqnarray}
Therefore, setting $z=E+{\mathrm i}0^+$, we find
\begin{eqnarray}
Re\Sigma_1(p,E+{\mathrm i}0^+)&=&
\sqrt{\frac{2\sigma^2}{\pi}}\,P\int_0^{\epsilon_{max}}\,
d\epsilon\,\frac{{\cal G}\left(q(\epsilon),p\right)}{E-\epsilon}\,,
\label{REGAUSS1L}\\
Im\Sigma_1(p,E+{\mathrm i}0^+)&=&-\sqrt{2\pi}\sigma^2{\cal G}(q(E),p)\,.
\label{IMGAUSS1L}
\end{eqnarray}
One can easily check that the imaginary part of the self-energy on the peak
is of order $p^{d+2}$, as previously announced. 

Turning now to the density of states, using Eq.(\ref{DOF}), 
Eqs.(\ref{REGAUSS1L}) and (\ref{IMGAUSS1L}), simplify now to
\begin{eqnarray}
Re\Sigma_1(\infty,E+{\mathrm i}0^+)&=&
\sqrt{\frac{2\sigma^2}{\pi}}\,P\int_0^{\epsilon_{max}}\,
d\epsilon\,\frac{q(\epsilon)\,{\mathrm e}^{-\sigma^2 q(\epsilon)^2/2}}{E-\epsilon}\,,
\label{REGAUSS1LDOF}\\
Im\Sigma_1(\infty,E+{\mathrm i}0^+)&=&-\sqrt{2\pi}\sigma^2
q(E)\,{\mathrm e}^{-\sigma^2 q^2(E)/2}.
\label{IMGAUSS1LDOF}
\end{eqnarray}
The zero$^{th}$ order approximation, 
$g_E(E)=\delta(E-\epsilon_{max})$, Eq.(\ref{EINSTEIN}), is largely 
modified at one loop. The imaginary part of the self-energy vanishes
at $0$ and $\epsilon_{max}$ since both ${\cal G}(q(0),p)$ and 
${\cal G}(q(\epsilon_{max}),p)$
are zero. The zero$^{th}$ order delta function is moved to a value 
$E^*$ which  verifies the relation~\cite{FOOTNOTEDELTA}:
\begin{equation}
E^*=\epsilon_{max}+Re \Sigma_1(\infty,E^*)\,,
\label{ESTAR}
\end{equation}
and the density of states at this order is 
\bq 
g_E(E)&=& 
\sqrt{\frac{2\sigma^2}{\pi}}
\frac{q(E)\,{\mathrm e}^{-\sigma^2 q(E)^2/2}}
{\left(E-\Sigma_1(\infty,E)\right)^2 + q(E)^2\,{\mathrm e}^{-\sigma^2 q(E)^2}}
\,,
 0\le E\le \epsilon_{max}\,, \\
g_E(E)&=&
\frac{1}{\left|\left.\frac{d \left[Re \Sigma_1(\infty,E)\right]}{d E}
\right|_{E=E^*}\right|}\delta(E-E^*)\,, 
E>\epsilon_{\max}\,. 
\label{DOSUNOCONT} 
\eq 

\subsection{Two loops}\label{TWOLOOPSECT}

The order 1/$\rho^2$ arises from including ${\cal I}^{(R)}_{R-2}(p)$
on the geometric series for the resolvent. The particle label repetitions
can basically arise in four ways, that we schematically depict
as
\begin{eqnarray}
&&\ldots 1\ldots 2\ldots 2\ldots 1\ldots\label{FACIL}\\
&&\ldots 1\ldots 2\ldots 1\ldots 2\ldots\label{DIFICIL}\\
&&\ldots 1\ldots 1\ldots 1\ldots\label{UNAPART}\\
&&\ldots 1\ldots 1\ldots 2\ldots 2\ldots\label{ALLADYSON}
\end{eqnarray}
The repetitions depicted in (\ref{FACIL}) and (\ref{DIFICIL}) are
genuine contributions to the self-energy. The repetitions
(\ref{ALLADYSON}) are contributions to the Dyson resumation of
$\Sigma_1$, (the $G^3_0(p,z)\Sigma^2_1(p,z)$ term in
Eq.(\ref{DYSONSUMA})), while the repetitions (\ref{UNAPART})
contribute both to $\Sigma_2$ and to $G^3_0(p,z)\Sigma^2_1(p,z)$.

In Appendix A we list all the two-loops terms  giving the
contribution to the self energy $\Sigma_2$.
Because of the complexity of the result, 
we shall focus herafter only onto the low momentum regime.
For $p=0$, there is a cancellation among all the various diagrams
and the self energy vanishes as it should.
The leading contribution to the
imaginary part of the self-energy scales as
\bq
Im \Sigma_2 (p,E) = C  p^2 q^{d-2}(E)
\label{IMASIGMA2}
\eq
In the low energy, low momentum region, close to the peak
$p \sim E^{1/2}$ of the structure factor,  it becomes:
\be
\Sigma_2 \propto p^2 E^{(d-2)/2} \propto p^d \ .
\label{EDI}
\ee
Consequently, for small values of the external momentum $p$,
the contribution to the width arising from the $1/\rho^2$ order
is  much  broader than the contribution from the $1/\rho$ order.
 In the frequency domain, the
expression (\ref{EDI}) gives indeed the broadening:
\be
\Gamma \propto p^{d-1}
\label{ALLELUIA}
\ee
i.e. $p^2$ in the three dimensional case, very different from 
the result (\ref{FREQUENZE}).

Putting all together, our final expression for the imaginary part
of the self-energy in the low momentum regime, for a function $f(r)=
\hat f (r/\sigma)$, is up to second order:

\be
Im \Sigma = \rf0 \left[ \frac{1}{\rho \sigma^d} (p
\sigma)^{d+2} \: A_1 + \left( \frac{1}{\rho \sigma^d} \right)^2
(p \sigma)^d \: A_2 \right]
\label{LOW}
\ee
where the coefficients $A_1$ and $A_2$ are pure numbers
 ${\cal O}(1)$ which can be determined 
after choosing the adimensional 
function $f(x)$ (the gaussian case will be studied in the
next section).
From the qualitative point of view it is worthwhile to note 
that (\ref{LOW}) describes the most general case.
It is quite easy to realize that even considering other multi-loops terms,
for a small momentum, the leading order term would be of the form 
(\ref{ALLELUIA}).

Let us note that the parameter of such expansions is actually the
inverse of the number of particles $N_R \equiv \rho \sigma^d$,
lying inside a cube whose dimensions are given by the range of the
interaction.
Let us picture the whole scenario as it emerges from two-loops
computation:

\begin{itemize}
\item
As a very general feature of topologically disordered systems, we obtain
the broadening $\propto p^{d-1}$ of the resonance peak of the
$S(p,\omega)$ at low enough momentum $p$, where the two-loop contributions
is dominant.
Because of the $p^{d-1}$ behaviour, topologically disordered systems
spread the energy of an incoming plane wave on a wider range of
frequencies than non-topologically disordered systems. This is
a quantitative characterization of the intuitive impression that  
they are ``more disordered''.
However, this feature is rather subtle, and only appears when 
the contributions $\propto 1/N^2_R$ to the self energy are considered. 

\item
At very high densities there exists a characteristic values $p_c$ where there is a
cross-over from the $p^{d-1}$ regime to the $\propto p^{d+1}$ regime.
From (\ref{LOW}) the cross-over momentum is easily found:

\be
(\sigma \: p_c)^2 = \frac{A2}{A1} \: \frac{1}{\rho \sigma^3}
\propto \frac{1}{N_R}
\label{CROSSOVER}
\ee

\item 
For $p>p_c$ the one loop term  mainly contributes and
the broadening is controlled by $\Sigma_1$. In this regime, the
asymptotic result (\ref{FREQUENZE}) has to be taken with a lot
of care. In fact it is supposed to be correct only for $\sigma p <<1$,
a regime which basically is dominated by the two-loops contributions. 
If $\sigma p_c$ is not small enough to be still in the same regime we are
no longer legitimate to assume that (\ref{FREQUENZE}) give the
correct results and we have to consider the full one-loop result
(\ref{SELF1}).
\end{itemize}
 
\section{Comparison with numerical simulations}\label{ONELOOPSECTNUM}

The dynamical structure factor $S_E(q,E)$ can be numerically computed. 
Its definition, Eq.(\ref{DEFINIZIONE}), implies that
it is a well normalized distribution function,
\begin{equation}
S_E(p,E)\ge 0\quad ,\quad \int_{-\infty}^{\infty}\, dE\, S_E(p,E) = 1\,.
\end{equation}
The moments of this distribution function can be calculated in terms
of the powers of the matrix $M$ defined in Eq.(\ref{EUCLIDEA}),
\begin{equation}
\int_{-\infty}^{\infty}\, dE\, S_E(p,E)\,E^R = \overline{
\frac{1}{N}\sum_{i,j}\,{\mathrm e}^{{\mathrm i} p (x_i-x_j)}\,
\left(M^R\right)_{i,j}
}\,.\label{LAEQMOM}
\end{equation}
If the function $f$ is non-negative, one can cut the previous
integrals at $E=0$, since the quadratic form associated to the matrix
$M$ is semi positive-definite:
\begin{equation}
\sum_{i,j} \varphi_i M_{i,j}\varphi_i = \frac{1}{2}\sum_{i,j} f(x_i-x_j)
(\varphi_i-\varphi_j)^2\ge 0
\end{equation}
Similarly, one can obtain the density of states replacing
Eq.(\ref{LAEQMOM}) by
\begin{equation}
\int_{-\infty}^{\infty}\, dE\, g_E(E)\,E^R = \overline{
\frac{1}{N}\sum_{i,j}\, \frac{v_iv_j}{\sum_k v_k^2}
\left(M^R\right)_{i,j}
}\, \ ,
\end{equation}
where the $v_i$ are random numbers chosen with uniform probability
between $-1$ and $1$, and the overline now means average over the
particle positions {\em and} the $v_i$.  Thus we see that one can uses
the method of moments~\cite{MOMENTI} complemented with a truncation
procedure~\cite{TRUNCAZIONE}.  Using the gaussian function $f$ of
Eq.(\ref{SCELTA}) truncated at four $\sigma$'s, we have been able to
reconstruct the $S_E(p,E)$ on systems with up to 32768 ($32^3$)
particles, using 100 moments. All the simulations in this paper have
been done generating ten samples of the disordered configurations: on
a box of side $L=32\sigma$, we place at random $\rho L^3$ particles.
We apply periodic boundary conditions, the minimum available momentum
thus being $\frac{2\pi}{32\sigma}$. Regarding the choice of the
density, let us recall that our natural length-unit is $\sigma$, that
can be considered as the analogous of the wavelength corresponding to
the first maximum of the dynamical structure factor. It is therefore
clear that the densities that correspond with the experimental
situation are of the order of $\sigma^{-3}$. In this paper we have
explored the range $0.2 \sigma^{-3}\le\rho\le 1 \sigma^{-3}$. In what
follows, momentum will be measured in units of $\sigma^{-1}$ and
density in units of $\sigma^{-3}$.

\begin{figure}[h!]
\begin{center}
\epsfig{file=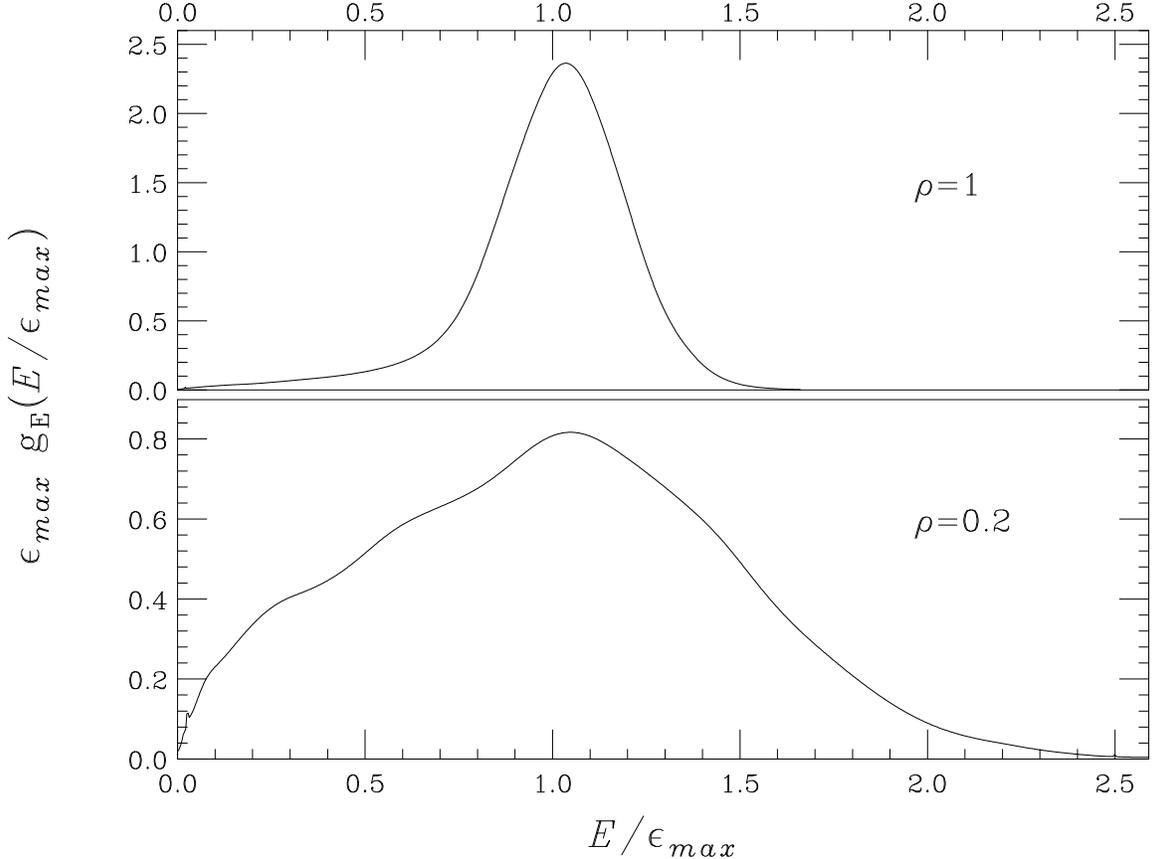,width=0.7 \linewidth,angle=90}
\end{center}
\caption{Density of states at densities $\rho=1$ and $0.2$. 
They have been obtained computing numerically $10$ moments (see text).
The X axis have been rescaled by $\epsilon_{max}$, and the Y axis have been also 
rescaled in order to have unit area under the curves.}
\label{COMPADOS}
\end{figure}

The choice of the number of calculated moments is conditioned by two
conflicting goals. On the first place, the irregularities on the
reconstructed probability distribution grows significantly with the
number of calculated momenta, which is due either to the statistical
fluctuations on the finite number of generated samples, or to round-off
errors on the Lanczos recursion. On the other hand using a too small
number of moments can smooth-out real features of the curves. As a
rule we show the curve reconstructed with the minimum number of
moments for which the width of the $S_E(p,E)$ is stable.  We should
emphasize that the moments method is a statistical one. With a finite
number of particles, there are only $\rho L^3$ available eigenvalues.
If the number of eigenvalues around the maximum of the $S_E(p,E)$ is
small, the method of moments does not yield meaningful
results. Moreover, since the $S_E(p,E)$ is an statistical quantity
itself, meaningful results cannot be obtained even with a full
diagonalization of the matrix $M$.  In order to get a feeling on what
difficulties might be encountered, it is useful to calculate the first
and second moments of the $S(p,E)$:
\begin{eqnarray}
\int_{-\infty}^{\infty}\, dE\, S_E(p,E)\,E &=& \ep\,,\\
\int_{-\infty}^{\infty}\, dE\, S_E(p,E)\,E^2 &=& \ep^2+\frac{2\rho}{(2\pi)^d}
\int dk \ff(k)[\ff(k)-\ff(p-k)]\,.\label{SDIPI2MOM}
\end{eqnarray}
It is clear in what sense  in the $\rho\to\infty$ limit the $S_E(p,E)$ 
becomes a
Dirac delta: its mean-value, $\ep$, grows like $\rho$, while its
variance only grows like $\sqrt{\rho}$. In particular, one can take
the $p\to\infty$ limit in the above equations, and see what happens
with the density of available eigenvalues. 
For the model described in Eq. (\ref{SCELTA}) , one
obtains:
\begin{eqnarray}
\int_{0}^{\infty}\, dE\, g_E(E)\,E &=&\rho (2\pi\sigma^2)^{3/2} \,,\\
\int_{0}^{\infty}\, dE\, g_E(E)\,E^2 &=&\rho^2 (2\pi\sigma^2)^3+
2\rho(\pi\sigma^2)^{3/2}\,.
\end{eqnarray}
The above results are rather discouraging, because one immediately
observes that at fixed $p$, the larger is $\rho$, the lesser states
are around $\ep$. Figure~\ref{COMPADOS} may help in clarifying 
this point.

The statistical sampling improves significantly with
decreasing densities, where however our analytical calculation is less
likely to work. In practice, the above considerations imply that the
smaller is $\rho$, the smaller is the minimum momentum for which the
$S_E(p,E)$ can be safely estimated on a finite box.

\begin{figure}
\begin{center}
\epsfig{file=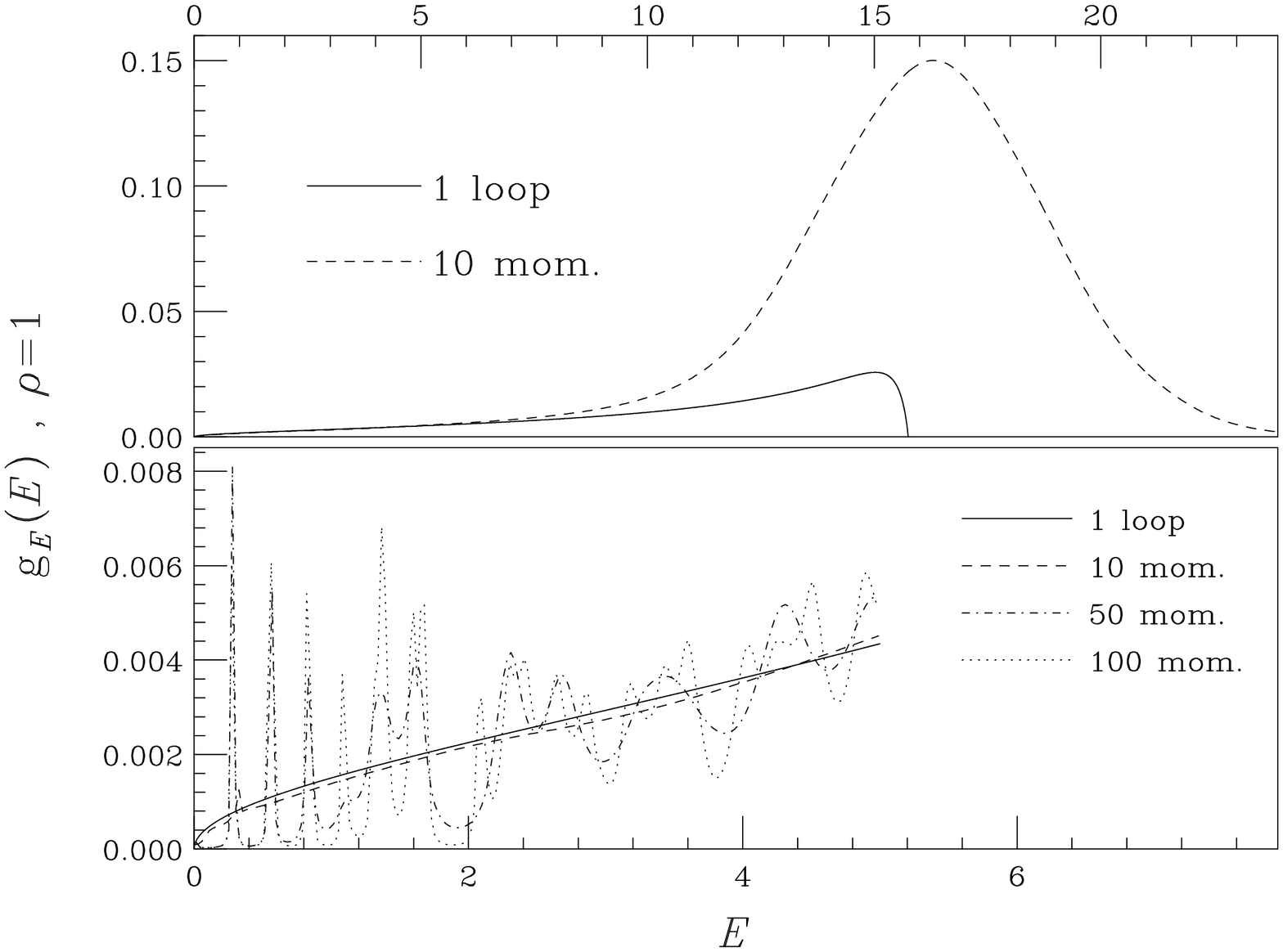,width=0.7 \linewidth,angle=90}
\end{center}
\caption{Numerical DOS compared with the analytical prediction at one-loop level 
(neglecting the contribution of the delta function).  The upper figure shows the 
comparison on the whole range of eigenvalues, and in the 
theoretical curve we have not 
drawn the delta function.  The lower figure focuses onto 
the low eigenvalues part of the 
spectrum, showing the dependence of the numerical data with the number of moments.}
\label{DOSONE}
\end{figure}

As a first comparison between the analytical and the numerical
computations, in figure (\ref{DOSONE}) we show the DOS obtained
numerically and the (continuous part) of the DOS at one-loop level as
obtained by means of the eq. (\ref{DOSUNOCONT})
 (neglecting the contribution of the delta 
function).  As it has been already
pointed out in~\cite{MePaZe}, the high density expansion is not really
suitable to describe the DOS in the neighborhood of the maximum.  In fact, we
have shown that at one loop the zeroth order delta function
splits from the continuous part of the DOS, carrying still a finite
weight.  Furthermore the continuous part (\ref{DOSUNOCONT})  is
defined only for $E<\eps_{max}$; in order to correctly compute 
the density of states in 
this region we need at least a partial resummation of the higher 
orders in the $\rho^{1}$ 
expansion, in the same way that it was done in ~\cite{MePaZe} for 
non translational 
invariant potentials. This task is not out of reach, however it 
goes beyond the scope of 
this paper. Because of this limitations, here only the low
eigenvalues region of the spectrum can be safely compared.  We see
that the numerical curve obtained with a large number of moments is
badly oscillating, because the single delta functions which form the
spectrum of a finite matrix are correctly reproduced by the method of
moments. If we consider the curve obtained considering only $10$
moments instead, there is a smoothing of the previous curves and we
see that it is in very nice agreement with the one loop computation.

Let us turn to the comparison of the dynamical structure factor. One
would like to check wether all the scaling regimes identified in 
subsection~\ref{TWOLOOPSECT} really hold.
One should therefore know at which value of the exchanged momentum
the one loop result will start to be dominating. The two coefficients
of Eq.(\ref{LOW}), for the Gaussian choice (\ref{SCELTA}), in the $3-d$ case 
the coefficients turn out to be:
\bq
A_1 &=& 7/12 \: \Omega_3 \\
A_2 &\approx& 1.25 \: \Omega^2_3 
\label{COEFFICIENTI} 
\eq 
with
\be
\Omega_3 = \int \: \frac{d \Omega_3}{(2 \pi)^3}
\label{ANGOLO}
\ee
The crossover momentum can then be estimated to be $0.31/\sigma$ for 
$\rho=1 \sigma^{-3}$ and $0.71/\sigma$ for $\rho=0.2 \sigma^{-3}$.
In a box of size $L=32 \sigma$ the first available momentum is $\sim
0.19/\sigma$, but unfortunately this is not the first momentum for which
one can reconstruct accurately the $S_E(p,E)$. Therefore, we should expect 
that our data can be accurately described using only by the full one-loop expressions.
Of course, one could try to compare with still lower densities, but this
would be both risky (ours is a $1/\rho$ expansion) and unphysical (glasses
have densities of order one, if measured in the natural length units of their
interaction potentials).

\begin{figure}[h!]
\begin{center}
\epsfig{file=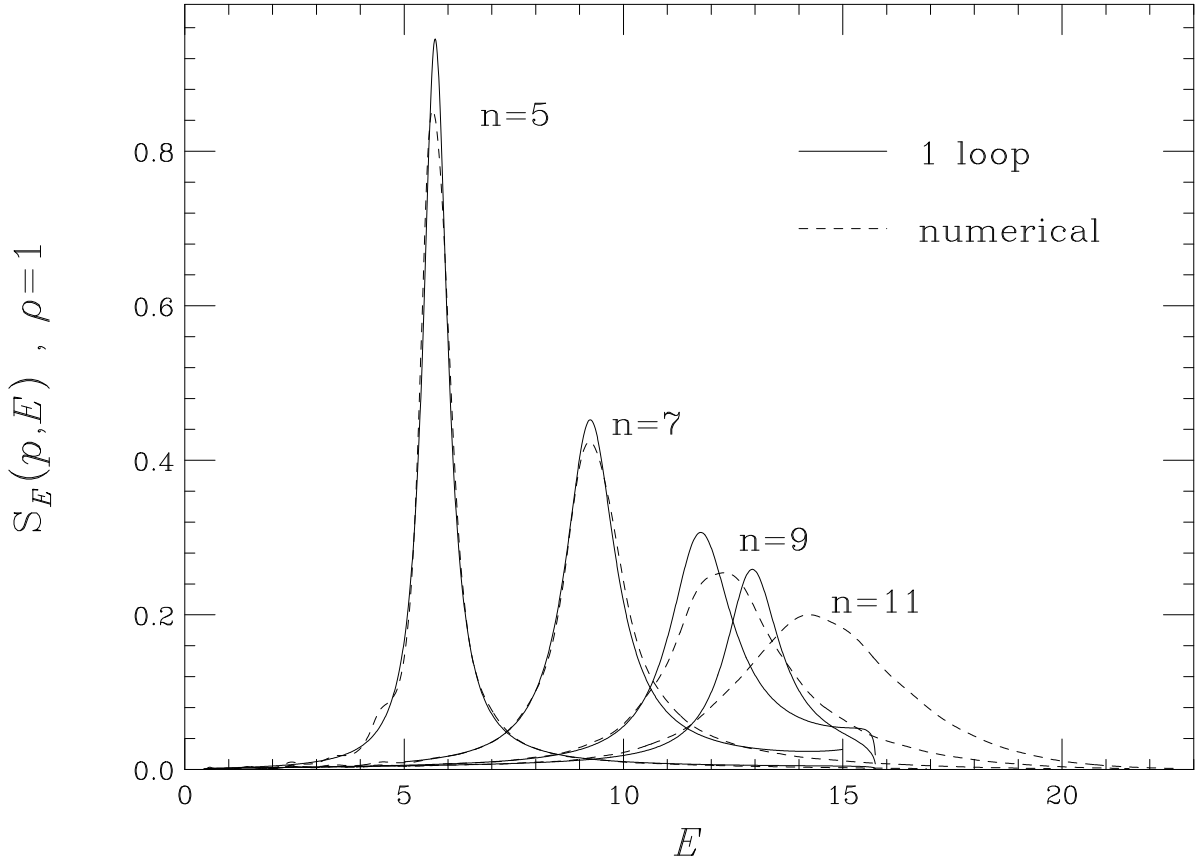,width=0.7 \linewidth,angle=90}
\end{center}
\caption{Dynamical structure factors at density $\rho=1$, the momentum
is given by $p = 0.19 n/\sigma$, with $n=5,7,9,11$. The numerical curves are
obtained with $30$ moments.}
\label{COMPAESSERHO1}
\end{figure}

The first comparison is for the density $\rho=1$. As  can be seen
from Fig.~\ref{DOSONE}, for a box with $32^3$ particles, when the
position of the peak is at $E_{peak} <1-2$ there are too few states
and the numerical results have no statistical significance. Hence, the
lowest value of the external momentum which is possible to compare
safely with the theory turns out to be $p=0.95/\sigma$ (i.e. $n=5$). The corresponding
results are displayed in Fig.~\ref{COMPAESSERHO1}.  We can see
that for $n=5,7$ the one-loop computation is in very good
agreement with the numerical data.  When the momentum is too large
($n=9,11$ in the figure), $E_{peak}$ becomes comparable to
$\eps_{max}$ and the agreement cannot be longer so good. The
analytical curve indeed is defined only up to $\eps_{max}$ (at all
orders in perturbation theory) and the tail which follows the peak
cannot be reproduced by the theory without further developments.

\begin{figure}[h!]
\begin{center}
\epsfig{file=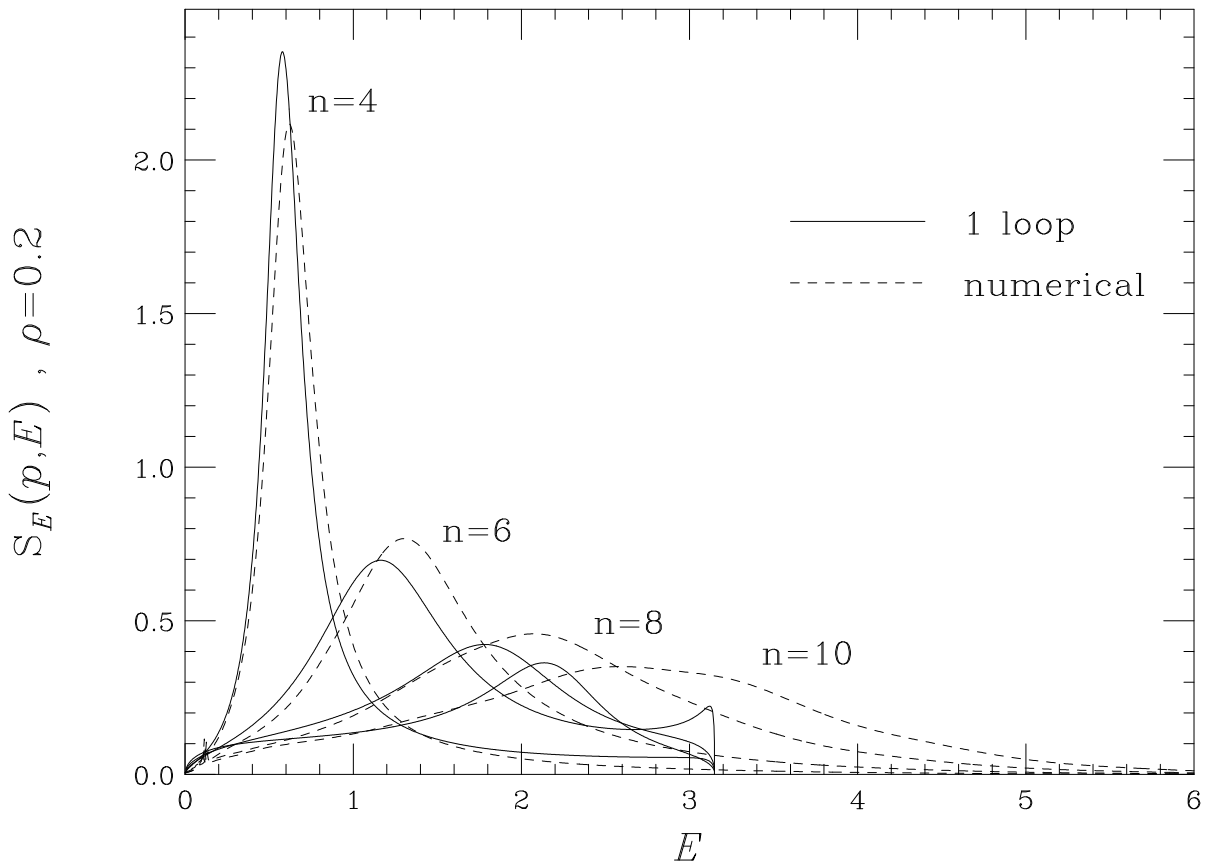,width=0.7 \linewidth,angle=90}
\end{center}
\caption{Dynamical structure factors at density $\rho=0.2$, the momentum
is given by $p = 0.19 n/\sigma$, with $n=4,6,8,10$. The numerical curves are
obtained with $10$ moments.}
\label{COMPAESSERHO0.2}
\end{figure}

The next comparison is at density $\rho=0.2$. Even if one might expect
that this density is too low for the $1/\rho$ expansion to be
accurate, the one-loop computation still describes quite well the
numerical curves, up to the region where $E_{peak}$ and $\eps_{max}$
become comparable.  One is tempted to conclude that the difference in
fig. (\ref{COMPAESSERHO0.2}) between theory and numerics for the
curves which are not strongly affected by the tail effect ($n=4,6$) are
due mainly to the following terms of the series in the $1/\rho$
expansion.  In particular the correction of the broadening is of order
$\sim\ 10\%$ while the position of the peak seems to be less affected by
higher order corrections, specially for the $n=4$ curve.

\begin{figure}[h!]
\begin{center}
\epsfig{file=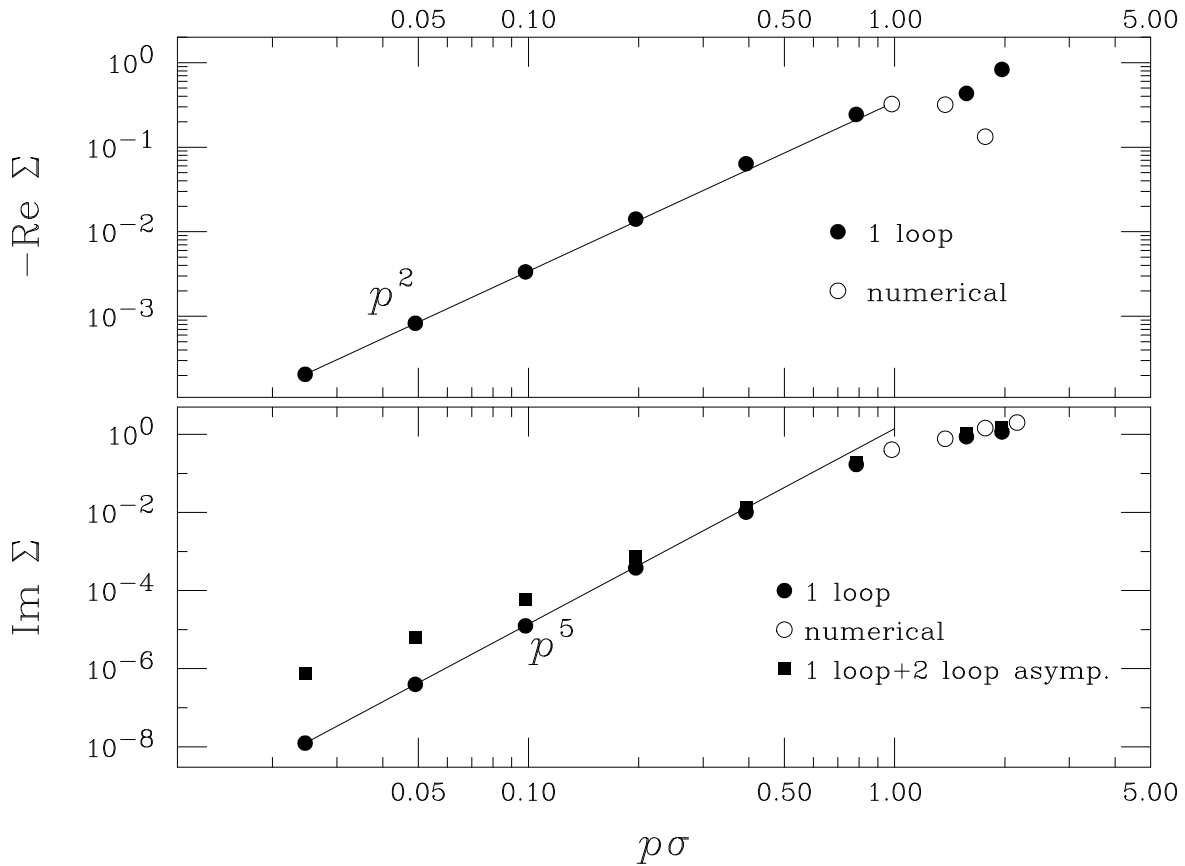,width=0.7 \linewidth,angle=90}
\end{center}
\caption{The real and the imaginary part of the self-energy 
computed at $E_{peak}$, for density $\rho=1$. 
The open circles are the numerical results, which cannot be extended
safely at lowest momenta. Full circles are the results of the
exact one loop computation. The squares contain  the
exact one loop computation corrected by the small $p$
expansion of the two loop result.}
\label{DISBRO}
\end{figure}

As we have pointed out in the introduction, 
a controversial point is the existence of a scaling law for the
broadening of the spectral peak with respect to $p$, in the range of
momenta explored by X-rays scattering experiments.
On the other hand, there is a general agreement on the existence of the
dispersion law for the peak position of the kind $E_{peak} \propto p^2$
($\omega^{peak}\propto p$),
up to momenta where the corresponding wavelength is comparable with the
mean separation of the particles, which enable us to identify the
peak as an acoustic one. Having shown that our numerical data are
reasonably described by the one loop computation, let us see
what are the predictions for these spectral features at this level
of accuracy. In the figure (\ref{DISBRO}) we plot
the real and the imaginary part of $\Sigma_1$ at the point 
$E_{peak}$, obtained by means of (\ref{REGAUSS1L},\ref{IMGAUSS1L}) 
as functions of $p$, for the density $\rho=1$
The real part follows the dispersion law $\propto p^2$ up to momentum
of order $1/\sigma$, and no significant dependence on the density is
observed while that scaling law still holds, see Eqs.(\ref{REASSI1L})
and (\ref{IMASSI1L}).  For the imaginary part, we already know from
the asymptotic analysis of the $p\to 0$ regime, that the one-loop low
momentum behaviour is $p^5$ ($p^4$ in the frequency domain), however
the range of momenta which are studied by numerical simulations, (and
by IXS scattering experiments!) surely are not in that asymptotic
regime.  The figure shows clearly that the lowest momentum that we are
able to numerically study is already in a region where that scaling
law does not hold anymore. As a matter of fact, the embarrassing
conclusion is reached that the scaling of the width of the dynamical
structure factor at one loop (which coincides with the numerical
simulation) could mimick a $p^2$ behaviour if the range of momenta
which is studied is  not  wide enough.

\section{About Particle Correlations}~\label{CORRELATIONS}

The formula (\ref{DIRETTA}) is  a good starting point to speculate
about the possibility of extending our computation to the case where
the particles are not chosen with a random distribution but, for
example, with the equilibrium Gibbs distribution. In that case, the
average on the position of the particles is made using  \cite{Ha}:
\be
\frac{1}{V^{R+1}} \int \prod^{R+1}_i d^d x_i g^{(R+1)}(x_1 \dots x_{R+1}),
\label{RANDOM2}
\ee
where $g^{(R+1)}$ is the $R+1$-points correlation function. Although
the computation using the full correlation function would be exceedingly
difficult, some progress can be made by using the so called superposition
approximation:
\be
g(x_1 \dots x_{R+1}) = g(x_1-x_2) g(x_2-x_3) \dots g(x_R-x_{R+1})
\label{SUPER}
\ee
where the pair correlation function is used to take into account the
correlation of the position of the particles.  The superposition
approximation has been probed to be reasonably correct in describing
the high order static correlation functions of the supercooled
liquids, both in the computation of the coupling coefficients in the
mode coupling theory \cite{GoMa} and in the computation of the
vibrational entropy \cite{MePa2,CoPaVe}.

It is not difficult to see that, to the lowest order in $1/\rho$, the
superposition approximation can be embebed in our calculation if we
make the substitution: 
\be f(r) \to \ g(r)f(r)
\label{FUTURO}
\ee 
This is rather important, because, for typical applications, the
function $f$ being badly divergent at low distances, does not a have a
Fourier transform. On the other hand, the function $g(r)$ typically 
tends to zero at the origin exponentially, thus taking care of the algebraic
divergence of $f(r)$.

However, the substitution (\ref{FUTURO}) cannot be blindly applied to
the calculation of the $1/\rho$ corrections. Let us consider for
instance the second moment (\ref{SDIPI2MOM}), that in the correlated case
yield (let $F$ be the Fourier transform of f(r)g(r)) 
\begin{eqnarray}
\hat\epsilon(p)&=&\rho F(0)-\rho F(p)\,,\\
\overline{M^2(p)}&=& [\hat\epsilon(p)]^2+ 
2\rho \int dx g(x) f^2(x)\left(1-e^{ipx}\right)\,
\end{eqnarray}
while the subsitution (\ref{FUTURO}) would have produced a factor
$g^2(x)f^2(x)$ in the above integral. A little thought reveals that,
for the $1/\rho$ calculation reported in Eq.(\ref{UNOSOBRERHOMOMENTO}), this
problem will arise whenever there is a string of only diagonal terms
($\delta_{k_i,k_{i+1}} f(x_{k_i}-x_{z_i})$) between the two repeated
particle labels. The diagonal terms are simply $[\rho F(0)]^b/z^{b+1}$
that add up to 
$$\frac{1}{z-\rho F(0)}\,.$$
Therefore, we conclude that the self-energy at order $1/\rho$ in the
superposition approximation is
\begin{eqnarray}
\Sigma_1(p,z)&=&\frac{1}{\rho}\int\frac{d q}{(2\pi)^D} 
[\rho F(p-q)-\rho F(q)]^2\left(\frac{1}{z-\hat \epsilon(q)}
-\frac{1}{z-\rho F(0)}\right)+\\\nonumber
&+& 2\rho \int dx g(x) f^2(x)\frac{\left(1-e^{ipx}\right)}{z-\rho F(0)}\,.
\end{eqnarray}
It is clear that none of the two new terms contribute to the imaginary
part close to $\hat\epsilon(p)$, and therefore our discussion of the
broadening of the spectral line, at this order of the $1/\rho$
expansion is basically unchanged. One could similarly discuss the
peculiarities of the $1/\rho^2$ term, but this is left for future work. 
However, the single term 
\begin{equation}
2\rho \int dx g(x) f^2(x)\frac{\left(1-e^{ipx}\right)}{z-\rho F(0)}\,,
\end{equation}
points to the possibility of a qualitatively new behaviour. 
Indeed, this term is the first
of a full category, in which the repeated particle labels always 
corresponds to the {\em same pair of particles}, separated by
a string of diagonal terms. It is not hard to show that the contribution 
to the self-energy of this family of terms is:
\begin{equation}
\frac{\rho}{2}\sum_{n=1}^\infty \int dx\, g(x) \left(1-e^{ipx}\right) 
\frac{[2f(x)]^{n+1}}{[z-\rho F(0)]^n}= 2\rho \int dx\, g(x) 
f^2(x)\frac{\left(1-e^{ipx}\right)}{z-\rho F(0)-2 f(x)}\,.
\end{equation}
This result implies that when the function $f(x)$ diverge at small
distance, the density of state will have a tail to infinity,
exponentially supressed by the $g(x)$ function. This exponential tail
is of a different origin from the instantonic contribution
recently  calculated in
Ref.\cite{ZEEAFFLECK}. Indeed the instantonic exponential tail arise
also for a fully regular f(x), due to anomalously dense regions.

\section{Discussion, Conclusions and Outlooks}\label{CONCLUSIONS}


The aim of this work was to study the spectral properties of euclidean
random matrices using an expansion on the inverse particle density
number. That approach is supposed to describe correctly topologically
disordered systems, in the context of the Instantaneous Normal Modes
approach to their high-frequency dynamics.  In particular, we have
performed a two-loop computation of the dynamical structure factor,
where the expansion parameter is the inverse of the density of
particles. The results have been analytically studied in the low
momentum regime, where the disorder reveals itself by the broadening
of the resonance peak.  The width of that spectral line has been
computed up to order $1/N^2_R$, where $N_R$ is the number of particles
which effectively interact with a given particle. The behaviour of the
width $\Gamma$ in the frequency domain, as function of the momentum
$p$ has been shown to follow a quite complicated law. When the
exchanged momentum, $p$, is quite smaller that the inverse wavelength
of the first maximum of the static structure factor, the width of the
spectral line follows $\Gamma \propto p^{2}$. This seem to be an
intrinsic difference between topologically-ordered and
topologically-disordered random systems, the Rayleigh $p^4$ scaling
being always found for the formers. At a momentum depending on the
particle density and the details of the potential, the $1/N_R$
contribution to the self-energy becomes dominant, and the $p^4$
scaling appears. Finally, when the exchanged momentum is of the order
of the inverse wavelength of the first maximum of the static structure
factor, the dynamical structure factor starts its collapse onto the
density of states. The width of the spectral line cannot grow at such
a violent pace as $p^{4}$, and if a not too large momentum-range is
explored, it can certainly mimick a $p^2$ scaling. It is possible that
the $p^2$ scaling found in the IXS experiments correspond to the later
transient behaviour.  Nevertheless, it would be very interesting to
explore experimentally a wider range of exchanged momentum (maybe
combining several scattering probes), in order to test this scaling
picture with three separated regimes.


It would be also very useful to compare the predictions of this computation
with the ones which have been obtained in the framework of mode coupling 
approximation \cite{GoMa}. 
There an hard sphere system is investigated and a $p^2$ law is obtained only 
for very low values of the frequency of the peak, which at the moment are 
not accessible to the experiments, while a crossover to a quite 
different behaviour is present when the external momentum moves to
higher values, comparable with the ones explored by the IXS scattering 
experiments. Even in that computation no clear scaling law actually emerges
and qualitatively that scenario is quite
similar to the one emerging from our approach.
A quantitative comparison though, is not straightforward because
our analytic computation does not concern any particular model 
of topologically disordered system, the choice (\ref{SCELTA}) 
being quite generic in order to understand which are the main features 
which determine the particular kind of broadening of the resonance peak.


The main point, however, is the reliability of our computation in 
understanding the experimental findings and, possibly, in predicting 
features not yet observed. 
Apparently, if one wants to apply the computation presented here
to the study of the spectral properties of realistic 
systems, such as silica or fragile glasses, one  faces the following 
limitations: 

\begin{itemize}
\item 
The computation has been explicitly performed in the case where particles are
uncorrelated which is deeply different from the situation found in
glasses and supercooled liquids.
\item
The values of the coefficients $A_1$ and $A_2$ in (\ref{LOW}) seems to depend
very badly on the choice of the interaction $f(r)$ hence at this level
is quite unclear the generality of the predictions that one could obtain.
from (\ref{CROSSOVER}) before performing the appropriate computations for a given system.
\end{itemize}
However, in section~\ref{CORRELATIONS} it has been shown that the
correlations between the particles can be at least partially taken
into account by performing the transformation (\ref{FUTURO}), which
amounts to {\em dress} the bare interaction $f(r)$ between two particles at
a given distance $r$, with the probability, given by the static
structure factor $g(r)$ of the real system to be studied, to find them
at that distance $r$.  Furthermore this procedure corresponds to the
well-known superposition approximation of the correlation functions of
the theory of liquids, a kind of approximation that has been largely
exploited previously in different analytical approaches to the study
of the glassy phase.  
More interestingly, the superposition
approximation makes even the latter limitation less severe.  As a
matter of fact the Gaussian choice (\ref{SCELTA}) has to mimic not
the bare interaction $f(r)$, that would be an hopeless task, but the
function $f(r) g(r)$ which sounds more feasible. We have also shown
how the terms for which the substitution $f(r)\to f(r) g(r)$ cannot be
made, give rise to an exponential tail of the density of states, of non
{\em instantonic} (see \cite{ZEEAFFLECK}) origin.
More in detail, by considering only the bare interaction one should take: 
\be f(r) = \frac{d^2 V(r)}{dr^2}
\label{LONGI}
\ee 
where $V(r)$ is the pair potential of the particles, which
typically has a strong repulsive barrier at $r \to 0$, missing in the
Gaussian choice.  The static structure factor $g(r)$, on the other
hand, is $\sim e^{-\beta V(r)}$ at $r \to 0$, hence the function $f(r)
g(r)$ is typically formed by a single peak in the position $r_0$,
where the static structure factor has its maximum, is not singular at
$r \to 0$ and it does have a finite Fourier transform looking quite
similar to (\ref{SCELTAQ}).  As a first approximation, the above
argument yields $\sigma \sim r_0$, i.e.  a 'few' Angstrom for
realistic systems, implying that X-rays experiments surely cannot
explore the low momentum region but they can observe the region $p \sigma > 0.2-0.3$, 
which is roughly the
same region as that investigated in the numerical simulations.

Hence we expect that the computation we have presented here, if
suitably modified for including the details of the system under study,
will allow to compute the dynamical structure factor of structural
glasses and supercooled liquids.

\section*{Acknowledgments}

We would like to thanks G.Biroli, W.Kob, R.Monasson, G.Ruocco and
G.Viliani for valuable discussions and crucial suggestions.  
This work has been performed in part during the stay of two of us
(MM and PV) at the Institute for Theoretical Physics
of the University of California, Santa Barbara. We thank the Institute
for its hospitality and acknowledge the partial support
 by National Science Foundation under grant
No.PHY94-07194. V.M.-M. is a M.E.C. fellow. P.V. would like also to
thank the SPHYNX program, which has supported his staying at the
Ecole Normale Superieure in Paris.

\newpage

\section*{APPENDIX A}

In this appendix we shall show the two loops integrals and the
corresponding diagrams arising from the $1/\rho$ expansion of the
resolvent to order 1/$\rho^2$.  It turns out to be useful to express
the integrals in terms of the propagators $\rff(q)$ and $G_s(q)$,
where \bq G_s(q) \equiv a^2 \left( G_0(q)-\frac{1}{a} \right) \nn \eq

\begin{figure}[h!]
\begin{center}
\epsfig{file=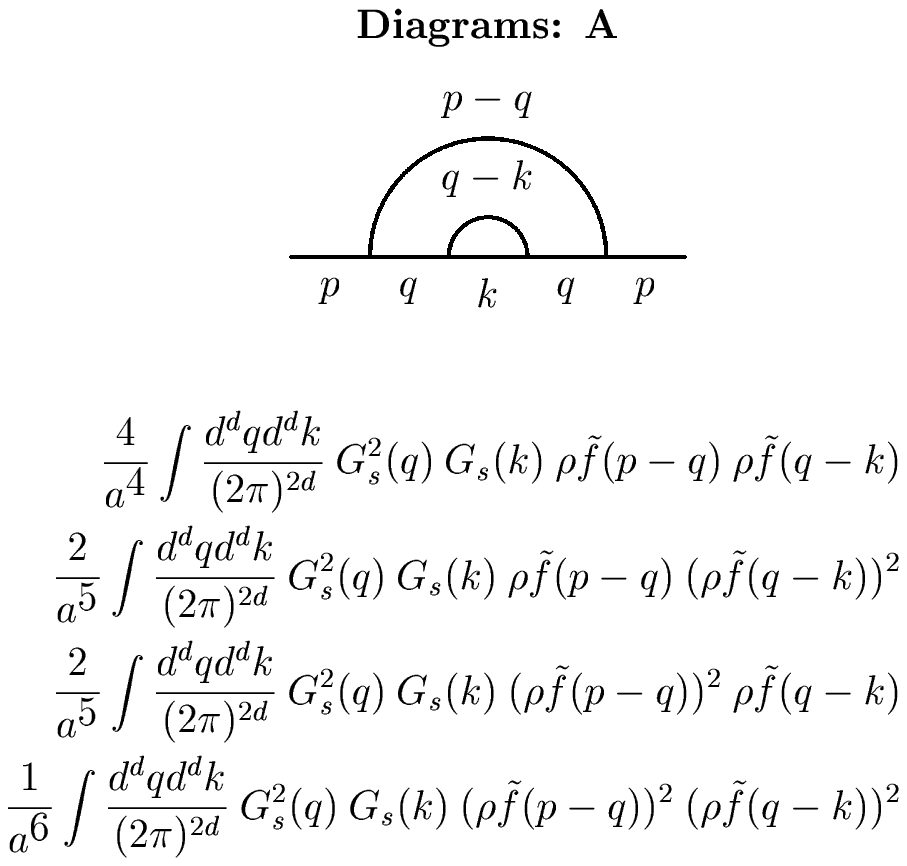,width=0.5\linewidth,angle=0}
\end{center}
\end{figure}

\begin{figure}[h!]
\begin{center}
\epsfig{file=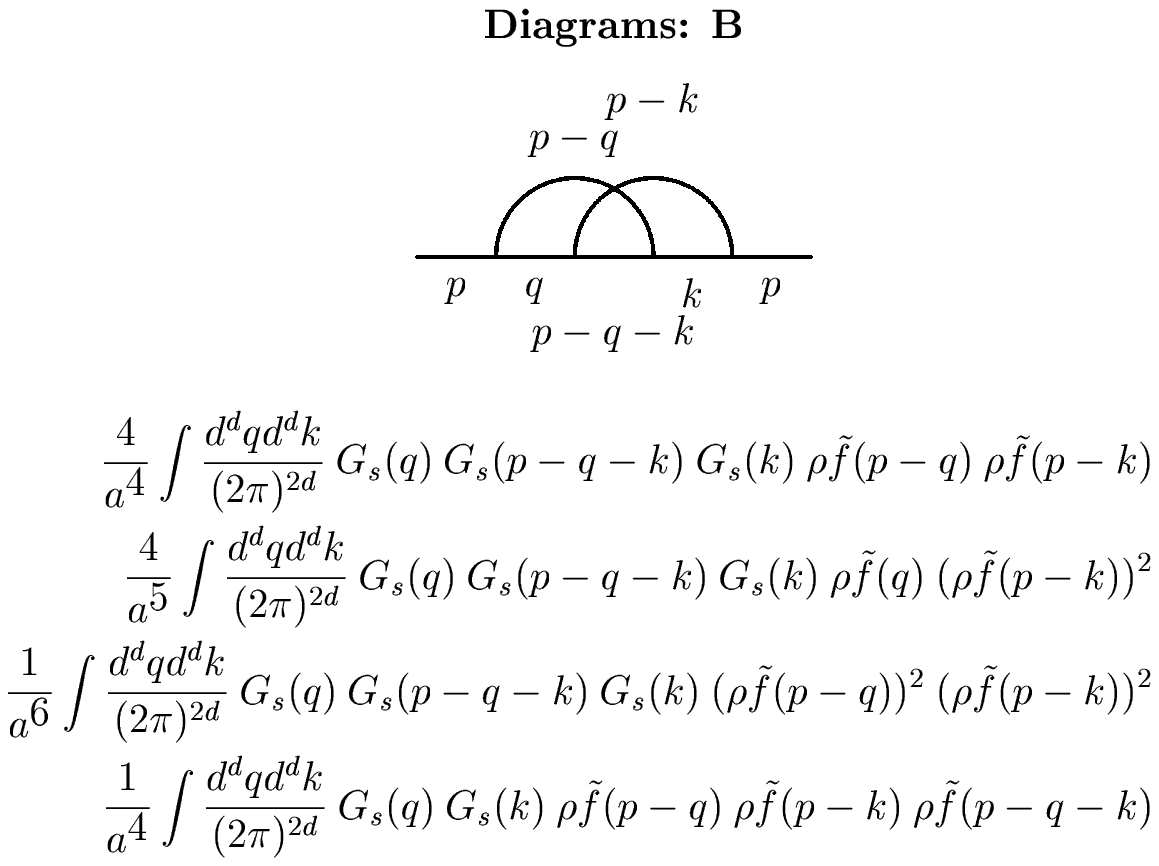,width=0.65\linewidth,angle=0}
\end{center}
\end{figure}

\begin{figure}[h!]
\begin{center}
\epsfig{file=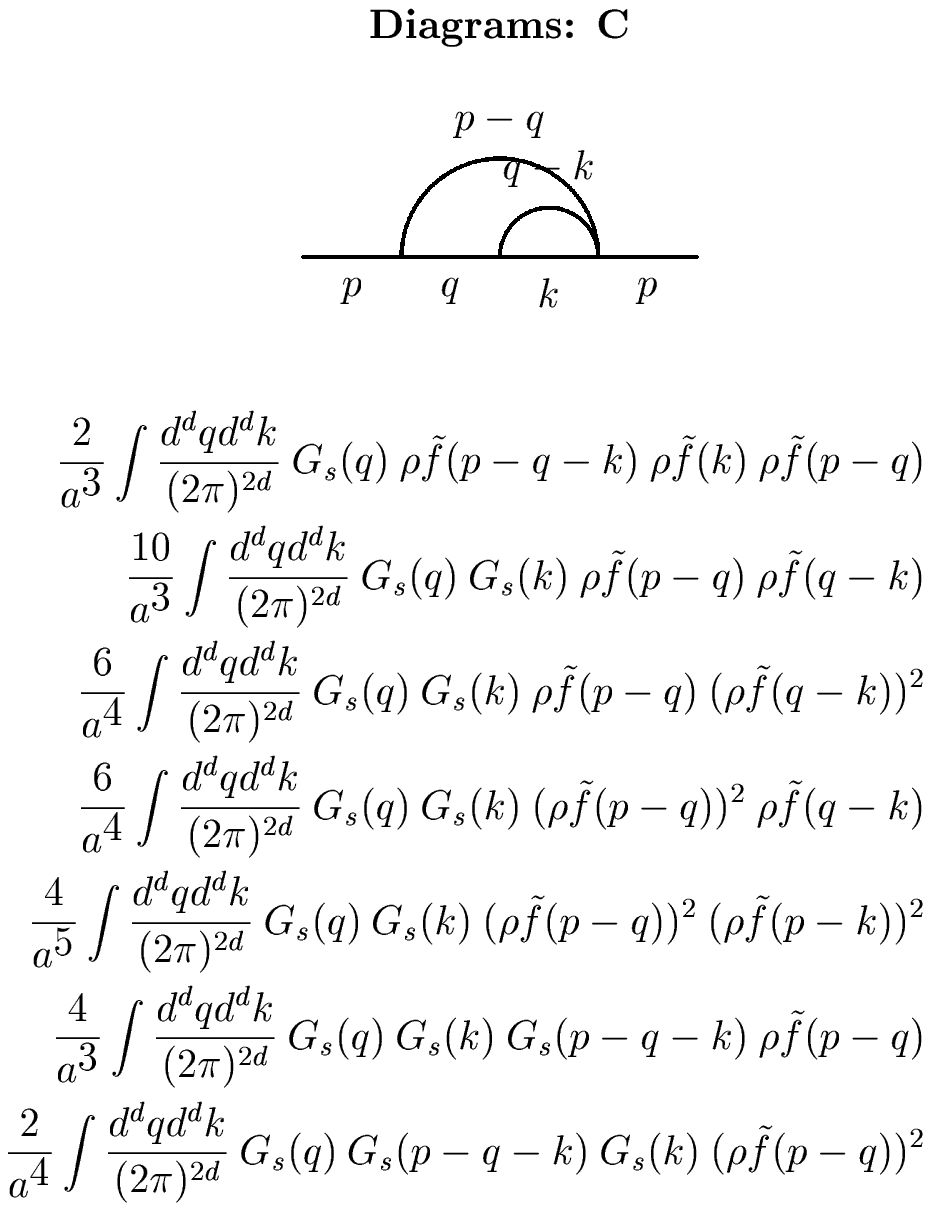,width=0.55\linewidth,angle=0}
\end{center}
\end{figure}

\begin{figure}[h!]
\begin{center}
\epsfig{file=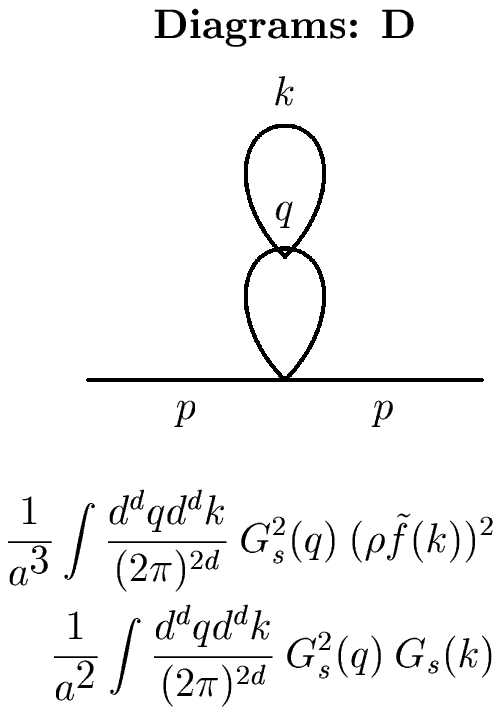,width=0.3\linewidth,angle=0}
\end{center}
\end{figure}

\begin{figure}[h!]
\begin{center}
\epsfig{file=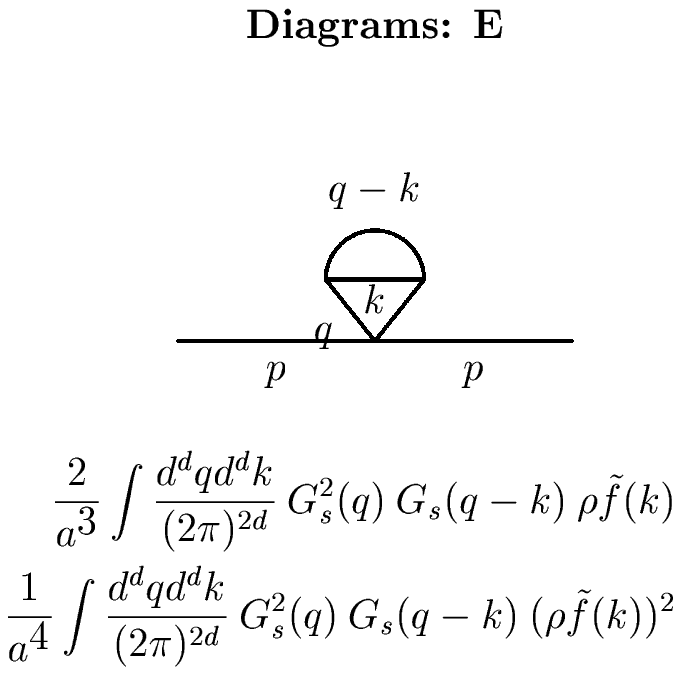,width=0.4\linewidth,angle=0}
\end{center}
\end{figure}

\begin{figure}[h!]
\begin{center}
\epsfig{file=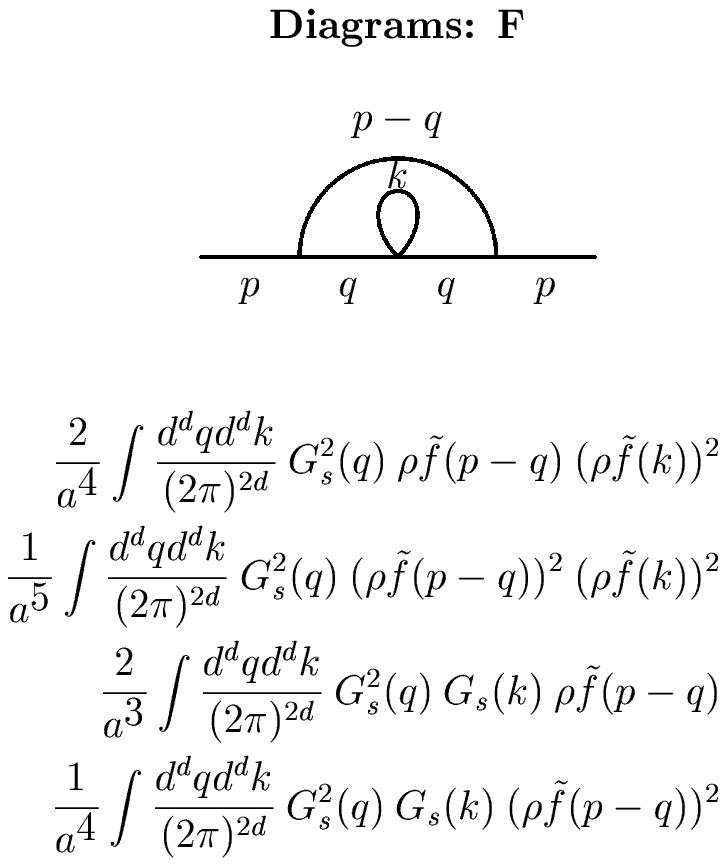,width=0.4\linewidth,angle=0}
\end{center}
\end{figure}

\begin{figure}[h!]
\begin{center}
\epsfig{file=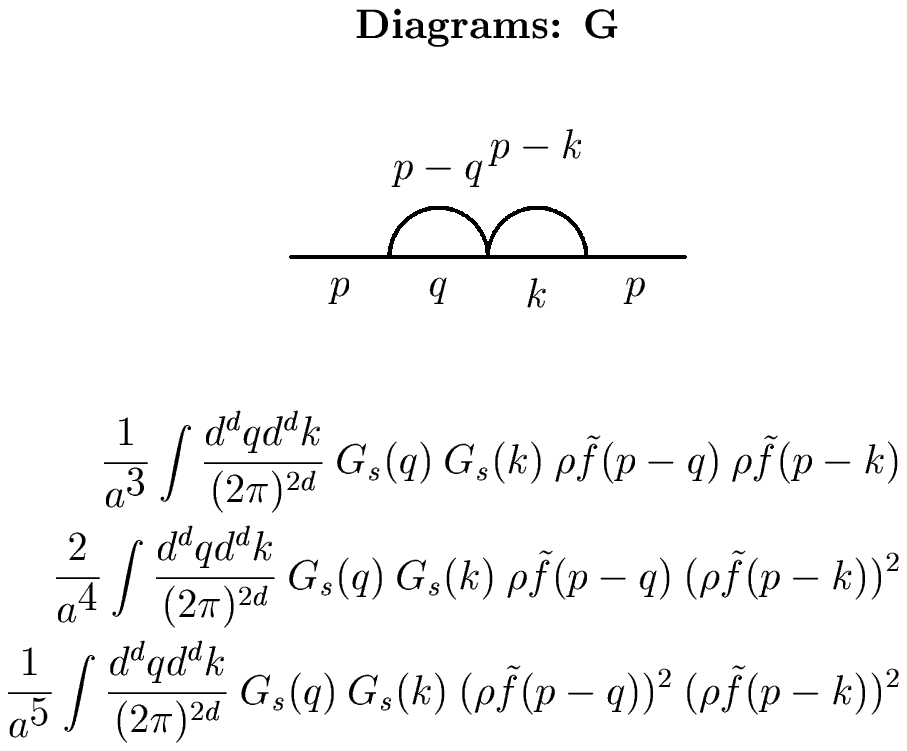,width=0.5\linewidth,angle=0}
\end{center}
\end{figure}

\begin{figure}[h!]
\begin{center}
\epsfig{file=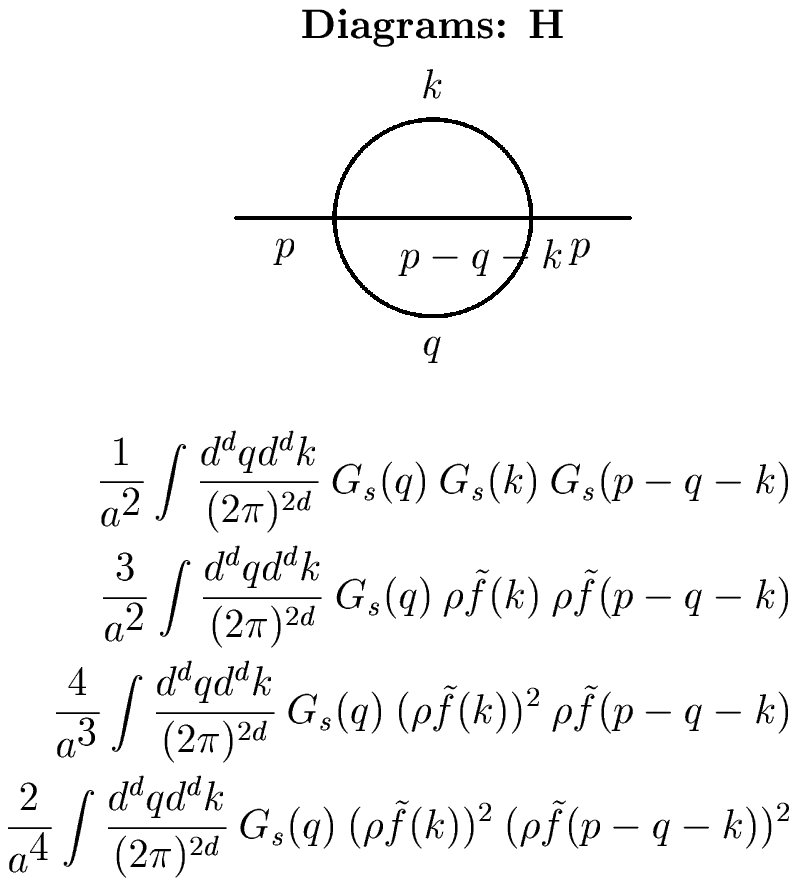,width=0.45\linewidth,angle=0}
\end{center}
\end{figure}

\begin{figure}[h!]
\begin{center}
\epsfig{file=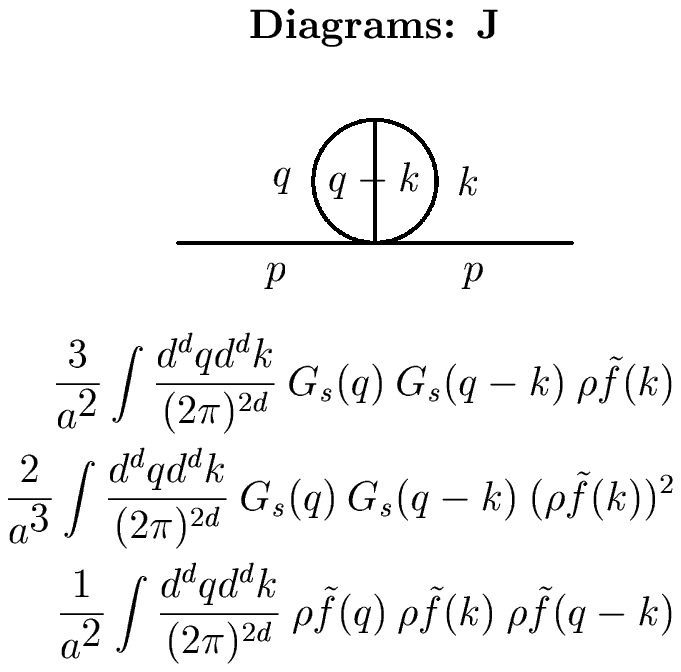,width=0.4\linewidth,angle=0}
\end{center}
\end{figure}

\begin{figure}[h!]
\begin{center}
\epsfig{file=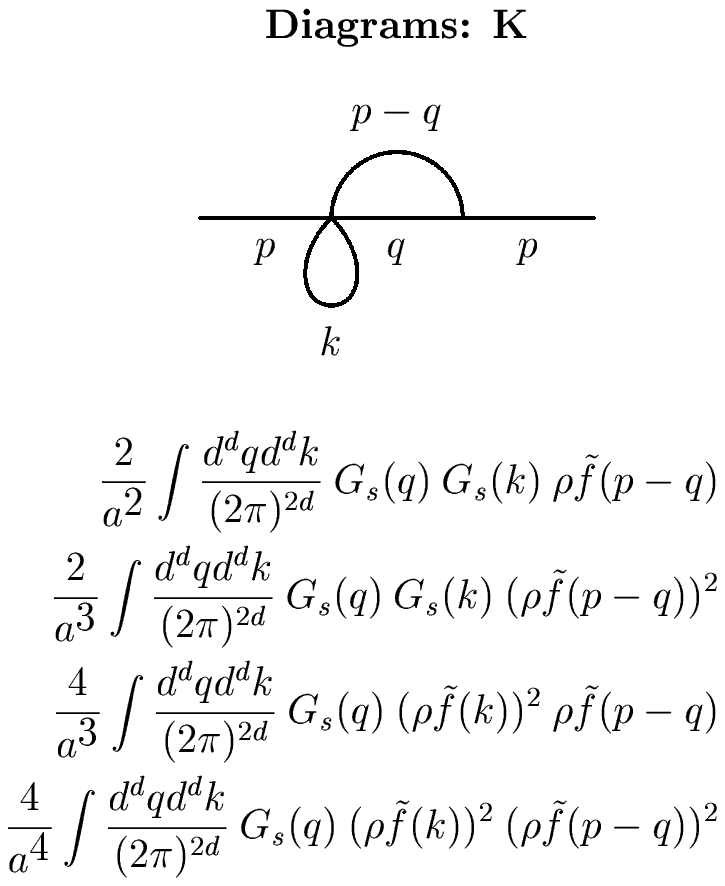,width=0.4\linewidth,angle=0}
\end{center}
\end{figure}

\begin{figure}[h!]
\begin{center}
\epsfig{file=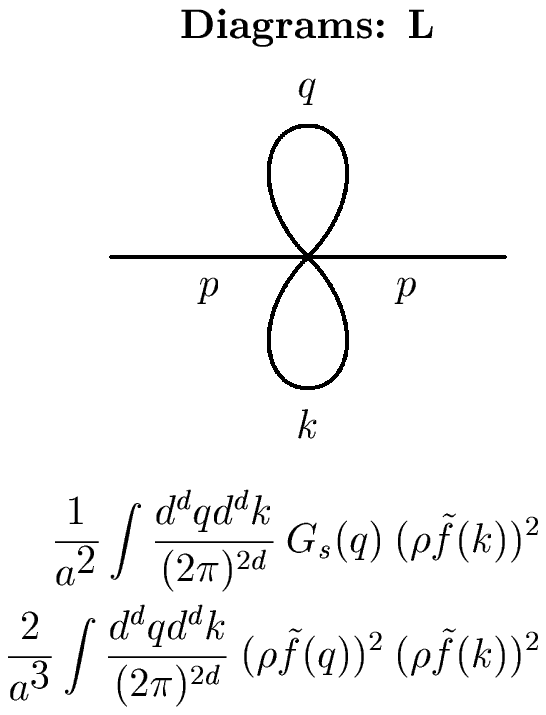,width=0.3 \linewidth,angle=0}
\end{center}
\end{figure}

\end{document}